# Quantum mysteries explained in digestible form.

**Alejandro A. Hnilo**

*CEILAP, Centro de Investigaciones en Láseres y Aplicaciones, (CITEDEF-CONICET);*
*J.B. de La Salle 4397, (1603) Villa Martelli, Argentina.*
*email: alex.hnilo@gmail.com*

*Abstract.*

Years ago, Itamar Pitowski asked two relevant questions: Why microphysical (quantum) phenomena and classical phenomena differ in the way they do? and, what kind of explanation could qualify as a reasonable one? I argue that both questions can be answered by the comparison of quantum phenomena with some features of vectors in real space. In particular, I show how violation of Bell's inequalities, Teleportation, Kochen-Specker and Greenberger-Horne-Zeilinger theorems can be understood in terms of vectors. This does not mean that the difference between quantum and classical phenomena is illusory. This means that vectors are stranger objects that they may seem to be at first sight.

## 1. Introduction.

Years ago, Prof. Itamar Pitowski put forward the following question, which he qualified as "the problem of interpretation" [1]: *"WHY is that microphysical phenomena and classical phenomena differ in the way they do?"* He also asked what kind of answer could qualify as a *reasonable* one.

Regarding the second question, in my opinion, an answer qualifies as "reasonable" when it uses ideas well-known in advance. In this way, reasonability of an answer is a relative feature. It depends on the ideas that are already well-known by the receiver of the answer. It would be then a situation similar to digestion: food (answer) is digestible (reasonable) only if the eater (receiver) has the appropriate enzymes (previous knowledge) to assimilate it. For a receiver with strong mathematical enzymes, stating that Quantum Mechanics (QM) algebra is an ortho-complemented non-distributive lattice may suffice to answer the first of Pitowski's questions. But for most people (including myself), such an answer just changes one mystery with another.

In this paper, I argue that the reason of the difference between microphysical (quantum) and classical phenomena is that the former requires a description in terms of non-Boolean entities (f.ex.: vectors), while the latter allows a description in terms of Boolean entities (sets). This should qualify as a reasonable answer to many, because vectors in real space, and sets, are mathematical objects learned in middle school. Besides, they are both visually understandable. It is only to be noted that, perhaps, one may have not noticed how alien to intuition the consequences of some vectors' features can be. In the next chapters I discuss the violation of Bell's inequalities (BI), Teleportation (or entanglement swapping, and the related Hong-Ou-Mandel effect), Kochen-Specker (KS) and Greenberger-Horne-Zeilinger (GHZ) theorems. I show how they can be understood by using vectors in real space.  I comment some relevant (and curious) vectors' features in the Appendix. I have discussed in detail the material in Chapters 2 and 3 before [2]; I review here the main results. Chapters 4 and 5 are fully new. I think it is useful to the Reader presenting all this material together.

Of course, the phenomena discussed here do not exhaust QM mysteries. Digesting the whole of QM is an enormous task that cannot be attempted in a single paper or by a single Author. Additional enzymes (as a minimum: complex algebra and vacuum fluctuations) are surely necessary. But, the mentioned phenomena are important enough to consider their explanation in digestible terms as a reasonable answer to (at least a significant part of) Pitowski's first question.

## 2. Violation of Bell's inequalities.

### *2.1 Stating the problem.*

I assume the Reader is familiar with the usual derivation of BI and basic mathematical analysis. Derivations of the Clauser-Horne inequality $J_{CH} \leq 0$, and the Clauser-Horne-Shimony and Holt one $S_{CHSH} \leq 2$ can be found in [3]. These derivations follow from the hypotheses of *Locality* and *Realism*. QM, on the other hand, predicts $J_{CH} \approx 0.207 > 0$ and $S_{CHSH} \approx 2.83 > 2$, in agreement with observations. The experimental verification of the violation of BI has led some researchers to declare physical reality independent of the observer ("Realism") to be demonstrated inexistent. The issue is important, because the existence of such reality is assumed not only in everyday's life, but also in all scientific practice. An alternative taken by many is to accept the existence of effects propagating at infinite velocity ("Non-Locality"), but this alternative is in conflict with the theory of Relativity. The conflict is usually circumvented by noting that these effects cannot be used to transmit information (what leads to theorems in Quantum Information theory dealing with "no signaling" restrictions). Yet, the tension between the two theories remains. That's why A.Shimony famously stated that QM and Relativity are in "peaceful coexistence", a term which had been coined to describe the fragile relationship between the East and West political and military blocks during the Cold War. This uncomfortable situation is often claimed to be caused by imperfect definitions of "Locality" and "Realism"; this issue is complex and subtle [4-6].

However, it is currently becoming clear that the cause of the conflict between BI and QM is the use of classical probabilities (at the "hidden variables" level) to derive the BI. When J.S.Bell derived his inequalities, he naturally used classical probabilities. After all, QM predictions are statistical. But classical probabilities presuppose (intuitive) Boolean logic [7]. Boolean logic also leads to Boole's *conditions of completeness*. G.Boole devised them, more than one century ago, as a criterion to decide whether a theory can be considered complete, or not. Boole's conditions, if applied to Bell's experiment (see Figure 1) are equivalent to the BI [1,8]. In consequence, for any complete Boolean theory, violating Bell's (or Boole's) inequalities is a logical impossibility. A theory with some hope to explain the violation of BI must start by giving up Boolean logic.

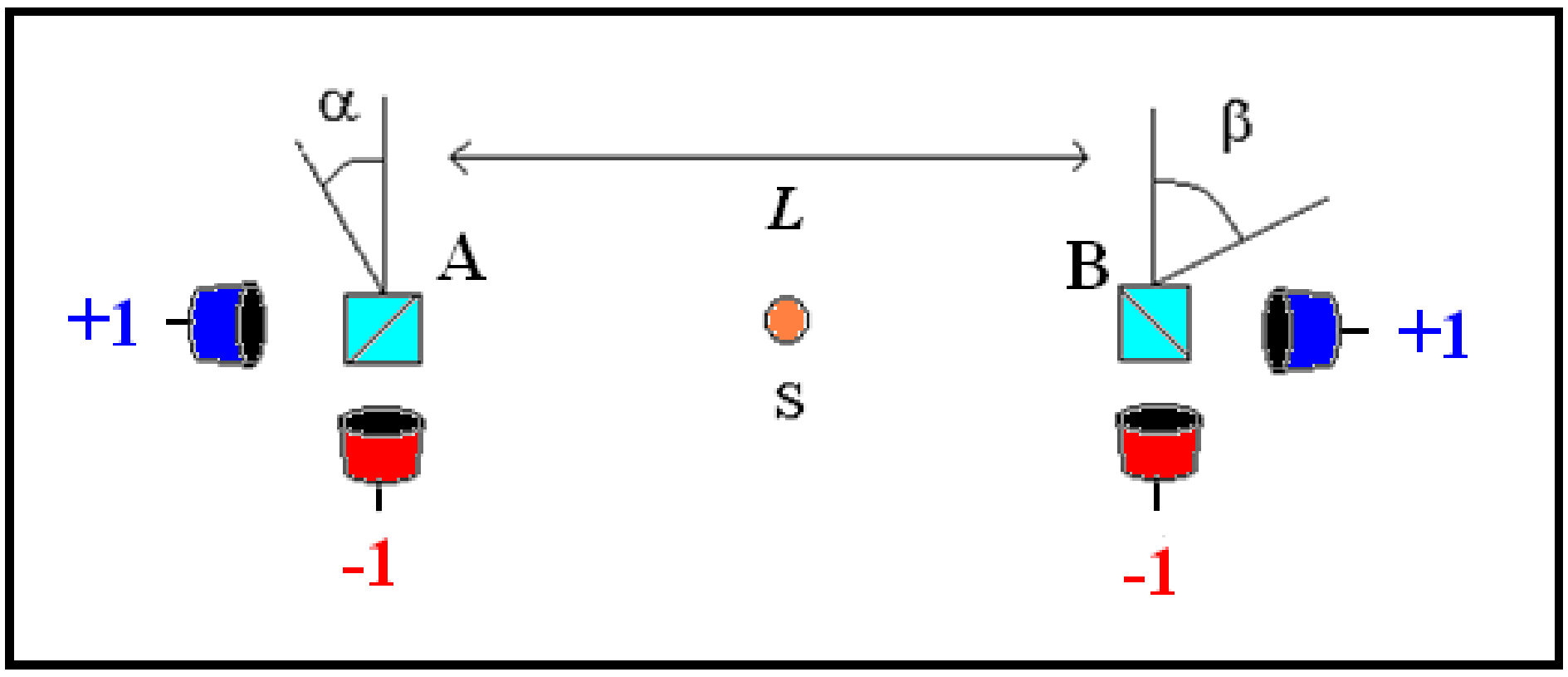

Figure 1: Sketch of a Bell's (or EPR-Böhm) experiment. Source **S** emits pairs of photons entangled in polarization, which propagate to stations A and B separated by a (large) distance *L*, and are detected after polarization analyzers oriented at angles {α,β}. The correlation between outcomes at each station violates classical bounds (Bell's and Boole's inequalities).

*2.2 A Boolean hidden variables model.*

The origin of the BI and their violation can be understood graphically by drawing a simple hidden variables (HV) *Boolean* model [9]. Let suppose that each photon in Fig.1 carries an angular HV named $\lambda$. As Boolean logic is assumed valid, classical probabilities can be assigned. Polarization analyzers are assumed to operate in the following way:

If $\lambda \in [\alpha-\pi/4, \alpha+\pi/4]$, then: $P^+(\lambda,\alpha) = 1$ and $P^-(\lambda,\alpha) = 0$, (2.1a)

If $\lambda \notin [\alpha-\pi/4, \alpha+\pi/4]$, then: $P^+(\lambda,\alpha) = 0$ and $P^-(\lambda,\alpha) = 1$, (2.1b)

where $P^+$ ($P^-$) is the probability to be transmitted (reflected), see Figure 2. The distribution $\rho(\lambda)$ of the HVs for a non-polarized source is defined uniform in $[0,\pi]$. The number of photons detected after a polarizer is then $N_\alpha = ½.N$ for all values of $\alpha$, as expected. Passage through two polarizers means "filtering" photons having features that allow passage through both polarizers. In Boolean logic, this corresponds to the operation "intersection" of the sets of features that allow passing through the polarizers. It does not matter if the polarizers are one after another, as in Fig.2, or in remote places as in Fig.1. The result of filtering is always given by the intersection of those two sets.

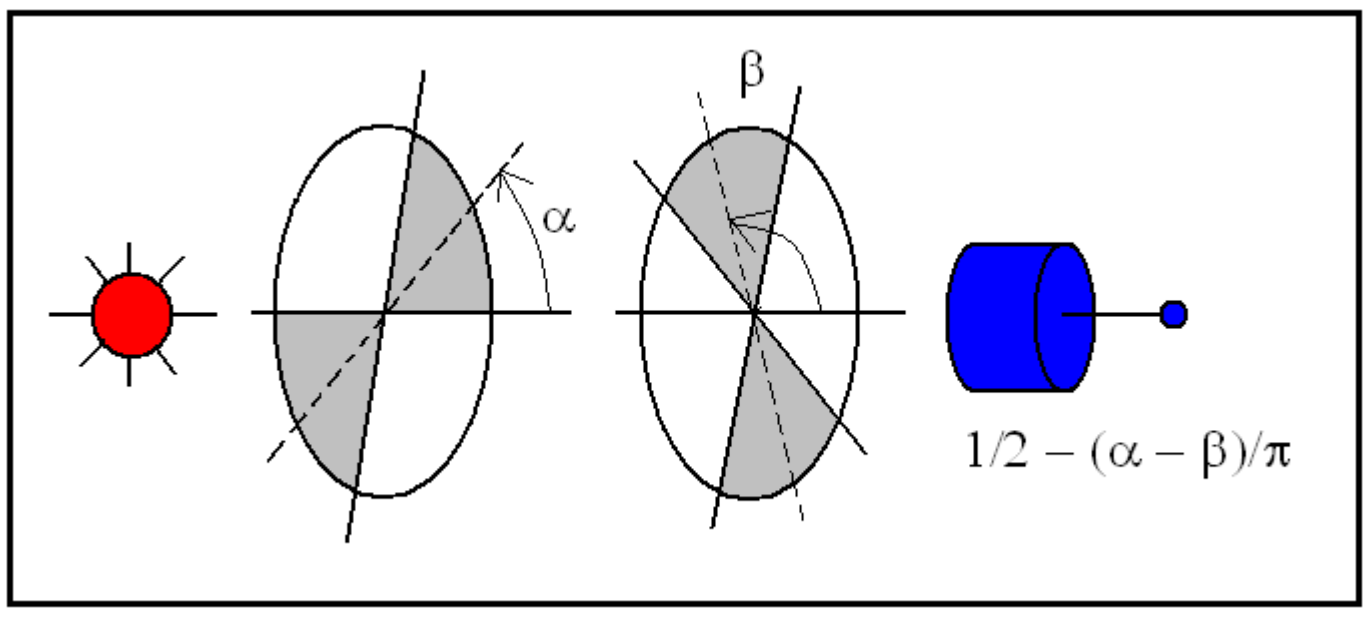

Figure 2: In the Boolean model, an analyzer can be thought of as a disk with two transparent (gray) and two reflecting (white) quadrants in the space of the hidden variable $\lambda$. Transmission through two analyzers is the saw tooth function: $½ - (\alpha-\beta)/\pi$, which is the "Boolean version" of Malus' law, regardless the analyzers are placed one after the other, as here, or in distant stations as in Fig.1.

Let suppose that the two photons in the pair carry the same value of the HV, $\lambda^A = \lambda^B$. The probability of coincidence is then given by a saw tooth curve:

$$P^{++}_{coinc,\ Boolean}(\alpha,\beta) = \int d\lambda.\rho(\lambda).P(\lambda,\alpha).P(\lambda,\beta) = [½ - (\alpha-\beta)/\pi] \qquad (2.2)$$

which is exactly at the boundary allowed by the BI, see Figure 3. For eq.2.2, $J_{CH} = 0$ and $S_{CHSH} = 2$. This simple Boolean model predicts perfect correlation and anti-correlation (and even fits the QM probability value at $\alpha-\beta=\pi/4$), but it fails to reproduce QM predictions and the experimental results at intermediate angles, as the angles used to violate the BI are ($\alpha-\beta=\pi/8$ and $3\pi/8$). In fact, the QM prediction for the (f.ex.) fully symmetrical Bell's state $|\varphi^+_{AB}\rangle = (|x_A\rangle\otimes|x_B\rangle+|y_A\rangle\otimes|y_B\rangle)/\sqrt{2}$ in the usual notation is:

$$P^{++}_{QM}(\alpha,\beta) = ½.cos^2(\alpha-\beta) \qquad (2.3)$$

which leads to $J_{CH} \approx 0.207 > 0$ and $S_{CHSH} \approx 2.83 > 2$, as stated before.

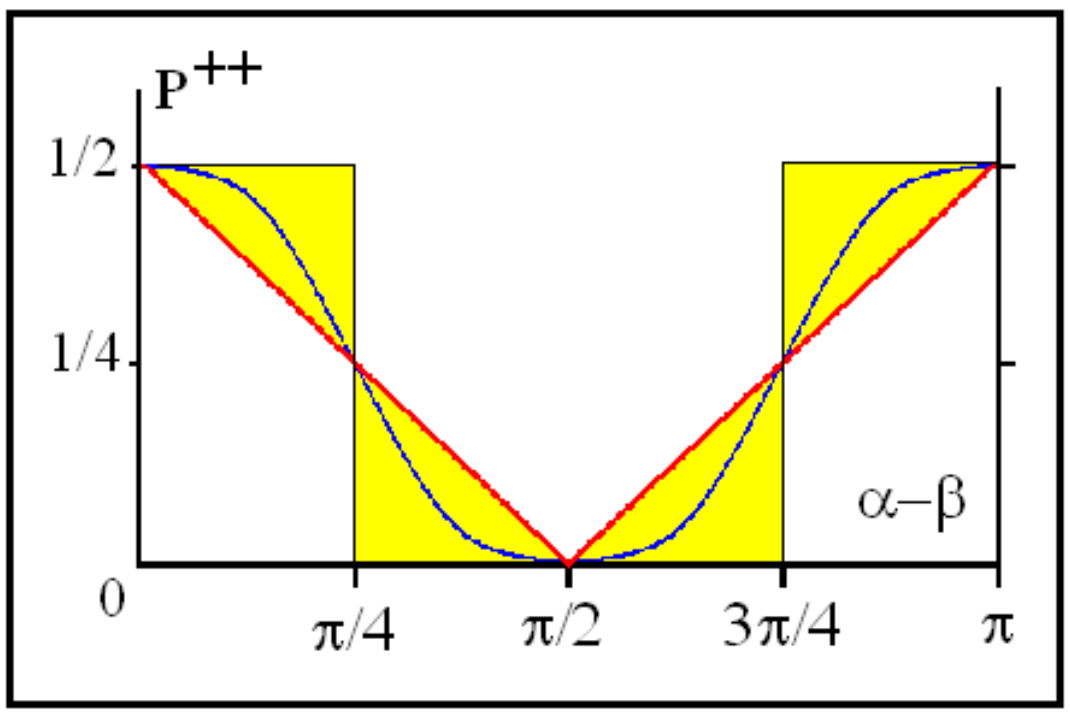

Figure 3: Prediction of the Boolean model for the experiment in Fig.1 is a saw tooth function (in red), which is also the limit allowed by the BI. QM, which algebra is non-Boolean, predicts a sinusoidal curve (blue, see eq.3) which is fully into the "forbidden" region (yellow). The difference between the red and blue lines is exaggerated here to enhance visibility.

*2.3 Violating Bell's inequalities with vectors in real space.*

As said, explaining the violation of BI needs a non-Boolean description. The simplest choice is a HV model using vectors in real space. Physical systems having a certain feature are now represented not by a set, but by a vector. Systems having more than one feature are not found as the intersection of the sets corresponding to those features, but as the projection of the corresponding vectors. In what follows, vectors are indicated with bold typing, scalars with normal typing. The operation "projection" of vector **b** into vector **a** is defined as:

$$\mathbf{a.b} \equiv |\mathrm{b}.cos(\gamma)|.\mathbf{e}_\mathrm{a} = (\mathbf{a}\times\mathbf{b}/\mathrm{a}).\mathbf{e}_\mathrm{a} \qquad (2.4)$$

where "**.**" (bold) means projection, "." means multiplication with a scalar, "×" means scalar product of vectors, $\gamma$ is the angle between **a** and **b**, a is the modulus of **a**, and $\mathbf{e}_\mathrm{a}$ is the unit vector in the direction of **a**. The resulting vector represents the systems that have features **b** and (then) **a** (be aware multiplication order is from right to left). F.ex.: exclusive features, which are represented by disjoint sets in Boolean logic, are now represented by orthogonal vectors. Note that the operation "projection" is neither commutative nor associative. It implies a non-Boolean algebra. The $cos(\gamma)$ factor in eq.2.4 anticipates why BI are violated (see Fig.3). Eq.2.4 looks much like the projection of state $|b\rangle$ into state $|a\rangle$ in QM: $|a\rangle\langle a|b\rangle$. But, be aware that the constituents of QM algebra are not simple vectors in real space, but closed vector subspaces or their corresponding orthogonal manifolds in a (complex) Hilbert space.

The main problem now is how to link vectors' modulus, which is a continuous variable, to discontinuous particle detection. A.Khrennikov has named this to be "the true quantum problem". Born's rule of QM is the simplest solution: the vector state's squared modulus gives the *probability* of detecting a particle. But, this solution fatally limits QM to be a statistical description. Besides, in the case of the experiment in Fig.1, it forces an entangled pair to be described as a single entity, say the Bell's state $|\varphi^+_{AB}\rangle$. This entity does not exist in real space, but in an abstract 4-dimensional *product* space of the spaces of two qubits. The state vector $|\varphi^{+}_{AB}\rangle$ has no internal parts, it is an "atom" (in the original ancient Greek meaning) of arbitrary large size. Not surprisingly, this strange entity seems to produce "nonlocal" effects (see also Chapter 5).

The next simple solution to relate vector length with particle detection is to define some threshold condition [10]. Suppose then that each photon in Fig.1 carries a real vector HV named **V**(t), which is orthogonal to the direction of propagation. The modulus of **V**(t) is V(t) (note bold typing for vectors); its unit vector is **v**(t), which is at an angle ν(t) with the *x*-axis; $\mathbf{V}(t) \equiv f(t).\mathbf{e}_x+g(t).\mathbf{e}_y$. It is tempting, and perhaps helpful, thinking **V**(t) as a physical field, say, the classical electric field. But it must be kept in mind that **V**(t) is a hypothetical HV. It can have the features that are found convenient as far as no contradiction arises. As all HV models, this one is neither intended to replace QM nor is claimed to be an accurate description of physical reality. It is intended just to make clear why there is no contradiction between the violation of BI and Locality or Realism (non Boolean versions, of course [2]).

In a (long) experimental run of duration *Tr*, the number $N$ >>1 of detected photons is:

$$N = (1/u).\int_0^{Tr} dt.\,|\mathbf{V}(t)|^2,\ N>>1 \text{ assumed} \tag{2.5}$$

where $u$ is a threshold value. The HV transmitted through a polarization analyzer oriented at an angle α are represented by projection of **V(t)** into the direction $\mathbf{e}_\alpha$. The detected number of particles $N_\alpha$ >>1 after the analyzer is then:

$$N_\alpha = (1/u).\int_0^{Tr} dt.\,|\mathbf{e}_\alpha.\mathbf{v}(t).V(t)|^2 = (1/u).\int_0^{Tr} dt.\,|\mathbf{e}_\alpha.cos[\nu(t)-\alpha].V(t)|^2 = (1/u).\int_0^{Tr} dt.\,V^2(t).cos^2[\nu(t)-\alpha] \tag{2.6}$$

In a non-polarized beam there is no correlation between ν(t) and V(t) nor with α (which is chosen by the observer), so the time average value of the cosinus-squared factor is ½:

$$N_\alpha = ½.N \tag{2.7}$$

for all values of α, as expected.

If the photons at A and B are entangled as $|\varphi^{+}_{AB}\rangle$, then their vector HVs are related as: $\mathbf{V^B}(t) = \mathbf{V^A}(t)$, thus $\nu^B(t) = \nu^A(t)$. Detections in station A are produced by the projection of $\mathbf{V^A}(t)$ into the axis $\mathbf{e}_\alpha$, i.e. $\mathbf{e}_\alpha.\mathbf{V^A}(t)$. These vectors exist in a counterfactual definite way (≈Realism). Hence, in the space of station B there exists a vector component of $\mathbf{V^B}(t)$ along direction α, $\mathbf{e}_\alpha.\mathbf{V^B}(t)$, which is identical to $\mathbf{e}_\alpha.\mathbf{V^A}(t)$, even if no observation is actually performed anywhere. The vector to calculate the number of coincidences $N^{++}$ at station B is not only $\mathbf{e}_\alpha.\mathbf{V^B}(t)$, but also the component in B-space corresponding to transmission through the B-polarizer. This feature, which implies further "filtering", is given by projection into $\mathbf{e}_\beta$. Therefore, by mere filtering in the space of the HVs (the same as in the case of the Boolean HV model) the vector component that produces *coincidences* $N^{++}$ in B-space is $\mathbf{e}_\beta.\mathbf{e}_\alpha.\mathbf{V}^\mathbf{B}(t)$. Then:

$$N^{++} = (1/u).\int_0^{Tr} dt.\,|\mathbf{e}_\beta.\mathbf{e}_\alpha.\mathbf{v}^\mathbf{B}(t).V^B(t)|^2 = (1/u).\int_0^{Tr} dt.\,|\mathbf{e}_\beta.\mathbf{e}_\alpha.cos[\nu^B(t)-\alpha].V^B(t)|^2 =$$

$$= cos^2(\alpha-\beta).(1/u.).\int_0^{Tr} dt.\,cos^2[\nu^B(t)-\alpha].V^2(t) = cos^2(\alpha-\beta).½.N \qquad (2.8)$$

Assuming as usual $P^{++} = N^{++}/N$, the QM prediction eq.2.3 is reproduced. The same result is obtained if the calculus is performed in station A. There are no entangled states here: $\mathbf{V}^\mathbf{A}(t)$ and $\mathbf{V}^\mathbf{B}(t)$ are vectors in real space, they are created at the source and then evolve separately. Note that this description use neither probabilities (it cannot, because defining probabilities presupposes Boolean logic) nor Born's rule, but a time integral and a threshold condition. Yet, as it is valid for long *Tr* and *N*>>1, its predictions are only statistical, as the QM ones are.

In short, the differences between the Boolean and the non-Boolean models are the nature of the HVs (sets and vectors) and the filtering operations involved (intersection and projection, see also the Appendix). In the Boolean HV model, the systems at stations A and B share a feature that is well defined since the moment of their emission: a Boolean HV named $\lambda$. Detection after an analyzer is determined by the condition that $\lambda$ belongs to a certain set. The number of coincidences is calculated by filtering, as the intersection of two sets: the one that corresponds to detection after the analyzer in A with the one that corresponds to detection after the analyzer in B, regardless the polarizers are one after the other or placed at distant stations. The operation intersection produces a function that varies *linearly* with the angle difference and is at the limit allowed by the BI.

In the non-Boolean vector-HV model, the systems at stations A and B share a feature that is well defined since the moment of their emission: a non-Boolean HV named $\mathbf{V}(t)$. The number of detections after an analyzer is determined by the vector component parallel to the analyzer's axis. The number of coincidences is calculated by filtering, as the time integral of the squared modulus of the vector component that results from two projections: of $\mathbf{V}(t)$ on the axis $\mathbf{e}_\alpha$ that corresponds to detection after the analyzer in A, and on the axis $\mathbf{e}_\beta$ that corresponds to detection after the analyzer in B, regardless the polarizers are one after the other or placed at distant stations. The operation "projection" produces a function that varies *quadratically* (at first order) with the angle difference, and hence violates the BI (see Fig.3).

This description using $\mathbf{V}(t)$ provides statistical results, thus it does explain the QM predictions. But it goes no further. It is not fully satisfactory, for it does not allow calculating *when* a single photon is detected (QM doesn't either). I have argued that such calculation can be done only in a non-Boolean logical framework [2]. A quantum computer (as it is currently known) is such a non-Boolean logical framework, but it uses Born's rule and is hence restricted to make statistical predictions only. A new type of quantum computer (or "wave computer", as D.Mermin described them) replacing Born's rule with a threshold condition is needed. Knowing **V**(t), this computer would be able to predict *when* a single photon is detected. Such new type of quantum computer does not exist yet, not even in theory. The task of imagining such computer appears formidable. Recall that the idea of "set" is fundamental in our usual definitions of "number" and "function". As Boolean logic and sets cannot be used for the purpose, new definitions without using the idea of "set" should be imagined. In the meantime, a classical computer (which is fatally Boolean) can *mimic* that new quantum computer if the *context* is defined [11] (see also the Appendix).

The vector-HV **V**(t), or something alike, would be the main "input" data to that new quantum computer. It seems reasonable thinking **V**(t) to represent phenomena in the atomic scale. Therefore, attempting to know **V**(t) in practice would be as irrational as attempting to know the position and momentum of all molecules in a mol of ideal gas. Hence, the new type of quantum computer would be relevant from the foundational point of view, but irrelevant for all practical purposes.

In summary: separable (Local) and counterfactual definite (holding to classical Realism) vectors in real space explain the violation of BI thanks to their non-Boolean algebra (filtering means *projection* instead of *intersection*). Therefore, the observed violation of BI disproves neither Locality nor Realism. It does disprove Boolean algebra at the quantum scale (what is, after all, hardly surprising).

## 3. Teleportation (or entanglement swapping).

### *3.1 Vector-Bell states.*

In the previous Chapter the relationship between the vector-HVs $\mathbf{V}^{\mathbf{A}}(t)$ and $\mathbf{V}^{\mathbf{B}}(t)$ was defined for the symmetry of the Bell state $|\varphi^{+}_{AB}\rangle$. For other Bell states, f.ex. $|\psi^{-}_{AB}\rangle = (1/\sqrt{2})\{|x_A,y_B\rangle-|y_A,x_B\rangle\}$, the relationship is $\nu_B(t) = \nu_A(t)-\pi/2$. This means that, for this Bell state, $\mathbf{V}^{\mathbf{B}}(t)$ is emitted (at the source) rotated with respect to $\mathbf{V}^{\mathbf{A}}(t)$. The first direction on which $\mathbf{V}^{\mathbf{B}}(t)$ is projected in eq.2.8 is not $\mathbf{e}_{\alpha}$ now, but $\mathbf{e}_{\alpha-\pi/2}$, and then $N^{++} = ½.N.cos^2[\alpha-\pi/2-\beta] = ½.N.sin^2(\alpha-\beta)$, as expected. Besides, if $\mathbf{V}^{\mathbf{A}}(t) = f(t).\mathbf{e}_x + g(t).\mathbf{e}_y$, then $\mathbf{V}^{\mathbf{B}}(t) = g(t).\mathbf{e}_x - f(t).\mathbf{e}_y$. The corresponding relationships for the remaining Bell states are:

for $|\psi^{+}\rangle$: $\nu_B(t) = -\nu_A(t)+\pi/2$, $\mathbf{V}^{\mathbf{B}}(t) = g(t).\mathbf{e}_{xB} + f(t).\mathbf{e}_{yB} \Rightarrow N^{++} = ½.N.sin^2(\alpha+\beta)$ (3.1a)

for $|\varphi^{-}\rangle$: $\nu_B(t) = -\nu_A(t)$, $\mathbf{V}^{\mathbf{B}}(t) = f(t).\mathbf{e}_{xB} - g(t).\mathbf{e}_{yB} \Rightarrow N^{++} = ½.N.cos^2(\alpha+\beta)$ (3.1b)

These results suggest the definition of (classical) *Vector-Bell states*:

$$\psi_{AB}^{-}(t) \equiv [f(t).\mathbf{e}_{xA} + g(t).\mathbf{e}_{yA}]\otimes[g(t).\mathbf{e}_{xB} - f(t).\mathbf{e}_{yB}] \quad (3.2a)$$

$$\psi_{AB}^{+}(t) \equiv [f(t).\mathbf{e}_{xA} + g(t).\mathbf{e}_{yA}]\otimes[g(t).\mathbf{e}_{xB} + f(t).\mathbf{e}_{yB}] \quad (3.2b)$$

$$\varphi_{AB}^{-}(t) \equiv [f(t).\mathbf{e}_{xA} + g(t).\mathbf{e}_{yA}]\otimes[f(t).\mathbf{e}_{xB} - g(t).\mathbf{e}_{yB}] \quad (3.2c)$$

$$\boldsymbol{\varphi}_{AB}^{+}(\mathrm{t}) \equiv [f(\mathrm{t}).\mathbf{e}_{xA} + g(\mathrm{t}).\mathbf{e}_{yA}]\otimes[f(\mathrm{t}).\mathbf{e}_{xB} + g(\mathrm{t}).\mathbf{e}_{yB}] \tag{3.2d}$$

Expanding eq.3.2.a (time dependence is omitted):

$$\boldsymbol{\psi}_{AB}^{-} = f.g.\mathbf{e}_{xA}\otimes\mathbf{e}_{xB} - f^2.\mathbf{e}_{xA}\otimes\mathbf{e}_{yB} + g^2.\mathbf{e}_{yA}\otimes\mathbf{e}_{xB} - g.f.\mathbf{e}_{yA}\otimes\mathbf{e}_{yB} \equiv (f.g, -f^2, g^2, -g.f) \tag{3.3}$$

Vector-Bell states can be normalized. F.ex. for $\boldsymbol{\psi}_{AB}^{-}$:

$$|\boldsymbol{\psi}_{AB}^{-}| = [\boldsymbol{\psi}_{AB}^{-}\times\boldsymbol{\psi}_{AB}^{-}]^{½} = \{(f.g)^2 + f^4 + g^4 + (g.f)^2\}^{½} = f^2 + g^2 = \mathrm{V}^2(\mathrm{t}) \tag{3.4a}$$

This result is the same for all vector-Bell states. The value of $\mathrm{V}^2(\mathrm{t})$ changes in time, but its average over the complete run is:

$$\langle \mathrm{V}^2(\mathrm{t})\rangle = (1/Tr).\int_0^{Tr} dt.\,\mathrm{V}^2(\mathrm{t}) = N.u/Tr = u/T \tag{3.4b}$$

where $T=Tr/N$ is the average time in which one particle is detected. Scaling $u/T=1$, all vector-Bell states are normalized. Note that $u/T$ can be interpreted as a "single particle average power", so that this scaling is physically reasonable.

Vector-Bell states are orthogonal among them. F.ex.:

$$\boldsymbol{\psi}_{AB}^{-} \times\boldsymbol{\varphi}_{AB}^{+} = (f.g, -f^2, g^2, -g.f)\times(f^2, f.g, f.g, g^2) = 0 \tag{3.5}$$

which holds for all time values. The same applies to $\boldsymbol{\psi}_{AB}^{+}\times\boldsymbol{\varphi}_{AB}^{-}$. For $\boldsymbol{\psi}_{AB}^{+}\times\boldsymbol{\varphi}_{AB}^{+}$ instead:

$$\boldsymbol{\psi}_{AB}^{+}\times\boldsymbol{\varphi}_{AB}^{+} = 2.\,f.g.[f^2+g^2] = 2.\mathrm{V}^2(\mathrm{t}).cos[\nu(\mathrm{t})].sin[\nu(\mathrm{t})] \tag{3.6a}$$

which, at a given time, is different from zero. Yet, its average over a complete run is zero:

$$\langle\boldsymbol{\psi}_{AB}^{+}(\mathrm{t})\times\boldsymbol{\varphi}_{AB}^{+}(\mathrm{t})\rangle = 2.\int_0^{Tr} dt.\,\mathrm{V}^2(\mathrm{t}).cos[\nu(\mathrm{t})].sin[\nu(\mathrm{t})] = 0 \tag{3.6b}$$

(recall that V(t) and ν(t) are independent). The same applies to $\boldsymbol{\psi}_{AB}^{-}(\mathrm{t})\times\boldsymbol{\varphi}_{AB}^{-}(\mathrm{t})$, $\boldsymbol{\psi}_{AB}^{+}(\mathrm{t})\times\boldsymbol{\psi}_{AB}^{-}(\mathrm{t})$ and $\boldsymbol{\varphi}_{AB}^{+}(\mathrm{t})\times\boldsymbol{\varphi}_{AB}^{-}(\mathrm{t})$.

In summary: vector-Bell states are normalized and orthogonal after averaging over a (long) observation time. Vector states corresponding to Eberhardt quantum states can also be written [2].

*3.2 Hong-Ou-Mandel effect.*

When two indistinguishable photons *A* and *B* enter a beam-splitter through each of its two input gates, they both leave on the same output gate ("bunching") unless they are in the Bell state $|\psi_{AB}^{-}\rangle$. This effect (Hong-Ou-Mandel) is used to experimentally project an undefined two photon state into $|\psi_{AB}^{-}\rangle$ and is crucial in the realization of teleportation or entanglement swapping. It is therefore pertinent, before going on, checking that this effect can be explained with vector-Bell states.

Let assume that the relationships between the vectors leaving a beam-splitter at the output gates *C,D* and the ones at the input gates *A,B* are the same than in classical Optics:

$$\mathbf{V}^C(\mathrm{t}) = [\mathbf{V}^A(\mathrm{t}) + \mathbf{V}^B(\mathrm{t})]/\surd 2; \qquad \mathbf{V}^D(\mathrm{t}) = [\mathbf{V}^A(\mathrm{t}) - \mathbf{V}^B(\mathrm{t})]/\surd 2 \tag{3.7}$$

Let define:

$$m^A{}_i = \int_{\theta_i}^{\theta_i+T} dt.|\mathbf{V}^A(t)|^2 = \int_{\theta_i}^{\theta_i+T} dt.[f^2(t) + g^2(t)] \tag{3.8}$$

where $\theta_i$ is an arbitrary time value, and $T$ is the time resolution of single photon detection (for an ideal detector, it is the photon bandwidth's inverse). An analogous expression applies for $m^B{}_i$. The *single incident photon condition*: $m^A{}_i = m^B{}_i = u$ is assumed.

Explicit time dependence is omitted in what follows. For vector-Bell state $\boldsymbol{\varphi}_{AB}{}^+$, $V_x{}^A = V_x{}^B = f$ and $V_y{}^A = V_y{}^B = g$. Using eq.3.7, $V_x{}^C = (f+f)/\sqrt{2}$, $V_y{}^C = (g+g)/\sqrt{2}$, $V_x{}^D = (f-f)/\sqrt{2}$, $V_y{}^D = (g-g)/\sqrt{2}$, then:

$$(V^C)^2 = (V_x{}^C)^2 + (V_y{}^C)^2 = (\sqrt{2}f)^2 + (\sqrt{2}g)^2 = 2.V^2(t) \tag{3.9a}$$

$$(V^D)^2 = (V_x{}^D)^2 + (V_y{}^D)^2 = 0 \tag{3.9b}$$

using eq.3.8 to integrate in time within each short interval, then:

$$m^C{}_i = 2.u\ ;\ m^D{}_i = 0 \tag{3.10}$$

this means that two photons are detected at one output gate and zero at the other, which is the expected bunching effect. For vector-Bell state $\boldsymbol{\psi}_{AB}{}^-$ instead:

$$(V^C)^2 = (V_x{}^C)^2 + (V_y{}^C)^2 = [(f+g)/\sqrt{2}]^2 + [(g-f)/\sqrt{2}]^2 = f^2 + g^2 \tag{3.11a}$$

$$(V^D)^2 = (V_x{}^D)^2 + (V_y{}^D)^2 = [(f-g)/\sqrt{2}]^2 + [(g+f)/\sqrt{2}]^2 = f^2 + g^2 \tag{3.11b}$$

so that, after integrating in time within each (short) interval $[\theta_i, \theta_i+T]$:

$$m^C{}_i = m^D{}_i = u \tag{3.12}$$

therefore, single photons are detected at both output gates in the same (short) time interval, as expected for the $\boldsymbol{\psi}_{AB}{}^-$ state.

The results derived from the vector-Bell states smoothly fit the QM ones until this point. The case of the two remaining vector-Bell states is a bit more involved, and is described in [2]. I mention here just that, in order to fully reproduce the bunching effect, an extra instruction is added to the way the beam splitter operates, stating (f.ex.) that the vector-HV with the smaller value of the time integral follows the path of the one with the bigger value. In this way, two photons are detected at one gate and zero at the other for $\boldsymbol{\varphi}_{AB}{}^-$ and $\boldsymbol{\psi}_{AB}{}^+$ too. If this extra instruction (say, "winner-take-all") is added, the QM predictions for the relationships between the *polarizations* at the two output gates [12], for all vector-Bell states, are reproduced too. Yet, even without the "winner-take-all" instruction, the only incoming vector-Bell state that produces simultaneous detections at the output gates is $\boldsymbol{\psi}_{AB}{}^-$, in agreement with QM predictions and experiments.

### *3.3 Teleportation in the QM and vector-Bell states cases.*

Teleportation in QM arises from an identity between the following combinations of quantum states:

$$|\psi_{12}{}^-\rangle\otimes|\psi_{34}{}^-\rangle = ½.\{|\psi_{14}{}^+\rangle\otimes|\psi_{23}{}^+\rangle - |\psi_{14}{}^-\rangle\otimes|\psi_{23}{}^-\rangle - |\varphi_{14}{}^+\rangle\otimes|\varphi_{23}{}^+\rangle + |\varphi_{14}{}^-\rangle\otimes|\varphi_{23}{}^-\rangle\} \tag{3.13}$$

As a consequence, if (f.ex.) systems 1 and 4 are measured together and found to be (thanks to the Hong-Ou-Mandel effect) in the state $|\psi_{14}^{-}\rangle$, then the systems 2 and 3, which have never interacted, are observed to be entangled in the state $|\psi_{23}^{-}\rangle$. Note that this occurs conditioned to the simultaneous detection of photons 1 and 4 after being combined in a beam splitter (post-selection). Also, that this is a wave phenomena, because of the condition of superposition in eq.3.13 and the orthogonality of Bell states.

Description in terms of vector-Bell states is straightforward and fully equivalent to the QM one. The corresponding identity among the vector-Bell states is (time dependence is omitted):

$$\boldsymbol{\psi_{12}}^{-}\otimes\boldsymbol{\psi_{34}}^{-} = ½.\{\boldsymbol{\psi_{14}}^{+}\otimes\boldsymbol{\psi_{23}}^{+} - \boldsymbol{\psi_{14}}^{-}\otimes\boldsymbol{\psi_{23}}^{-} + \boldsymbol{\varphi_{14}}^{+}\otimes\boldsymbol{\varphi_{23}}^{+} + \boldsymbol{\varphi_{14}}^{-}\otimes\boldsymbol{\varphi_{23}}^{-}\} \qquad (3.14)$$

which is valid for all time values, not only in the average over time *Tr*. The only difference between eq.3.13 and 3.14 is one term (the $\varphi^{+}$'s) with the opposite sign. It means a phase difference with no observable consequences in a single entanglement swapping process. It has been shown that QM predictions in some entanglement swapping network scenarios can be reproduced by using complex numbers only [13]. Here, vector-Bell states are defined real for simplicity, but they can be defined complex as well. This issue remains to be explored.

Note that the terms in eq.3.14 are consistent:

$$\boldsymbol{\psi_{14}}^{+}\otimes\boldsymbol{\psi_{23}}^{+} = (f.\mathbf{e}_{x1}+g.\mathbf{e}_{y1})\otimes(g.\mathbf{e}_{x4}+f.\mathbf{e}_{y4})\otimes(g.\mathbf{e}_{x2}-f.\mathbf{e}_{y2})\otimes(-f.\mathbf{e}_{x3}+g.\mathbf{e}_{y3}) \qquad (3.15a)$$

$$\boldsymbol{\psi_{14}}^{-}\otimes\boldsymbol{\psi_{23}}^{-} = (f.\mathbf{e}_{x1}+g.\mathbf{e}_{y1})\otimes(g.\mathbf{e}_{x4}-f.\mathbf{e}_{y4})\otimes(g.\mathbf{e}_{x2}-f.\mathbf{e}_{y2})\otimes(-f.\mathbf{e}_{x3}-g.\mathbf{e}_{y3}) \qquad (3.15b)$$

$$\boldsymbol{\varphi_{14}}^{+}\otimes\boldsymbol{\varphi_{23}}^{+} = (f.\mathbf{e}_{x1}+g.\mathbf{e}_{y1})\otimes(f.\mathbf{e}_{x4}+g.\mathbf{e}_{y4})\otimes(g.\mathbf{e}_{x2}-f.\mathbf{e}_{y2})\otimes(\mathrm{g}.\mathbf{e}_{x3}-f.\mathbf{e}_{y3}) \qquad (3.15c)$$

$$\boldsymbol{\varphi_{14}}^{-}\otimes\boldsymbol{\varphi_{23}}^{-} = (f.\mathbf{e}_{x1}+g.\mathbf{e}_{y1})\otimes(f.\mathbf{e}_{x4}-g.\mathbf{e}_{y4})\otimes(g.\mathbf{e}_{x2}-f.\mathbf{e}_{y2})\otimes(\mathrm{g}.\mathbf{e}_{x3}+f.\mathbf{e}_{y3}) \qquad (3.15d)$$

In other words: consider f.ex. that the vector for photon #1, $\mathbf{V_1}$(t), is arbitrarily chosen to be $f.\mathbf{e}_{x1}+g.\mathbf{e}_{y1}$. Then $\mathbf{V_2}$(t) must be $g.\mathbf{e}_{x2}-f.\mathbf{e}_{y2}$ because photons #1 and #2 are emitted in the same physical process (say, frequency down conversion in a type-II crystal), which produces a $\boldsymbol{\psi}^{-}$ vector-Bell state. As the form in eq.3.15a is a $\boldsymbol{\psi_{14}}^{+}$ state, then $\mathbf{V_4}$(t) = $(g.\mathbf{e}_{x4}+f.\mathbf{e}_{y4})$; and $\mathbf{V_3}$(t) = $(-f.\mathbf{e}_{x3}+g.\mathbf{e}_{y3})$ to get a $\boldsymbol{\psi_{23}}^{+}$ state. This expression for $\mathbf{V_3}$(t) is also the one necessary to fit the $\boldsymbol{\psi_{34}}^{-}$ state produced in the actual physical process of emission in the type-II crystal. The same applies to the other three equations. Finally:

$$\boldsymbol{\psi_{12}}^{-}\otimes\boldsymbol{\psi_{34}}^{-} = (f.\mathbf{e}_{x1}+g.\mathbf{e}_{y1})\otimes(g.\mathbf{e}_{x2}-f.\mathbf{e}_{y2})\otimes(f.\mathbf{e}_{x3}+g.\mathbf{e}_{y3})\otimes(-g.\mathbf{e}_{x4}+f.\mathbf{e}_{y4}) \qquad (3.16)$$

which fits the physical process of emission of pairs 1-2 and 3-4, as expected.

Let suppose then that two entangled pairs of systems are independently created: 1 and 2, and 3 and 4, with their vector-HV related as the vector-Bell states $\boldsymbol{\psi_{12}}^{-}$ and $\boldsymbol{\psi_{34}}^{-}$. Now let suppose that systems 1 and 4 are combined in a beam-splitter. If detections occur at both output gates in the time interval $[\theta_i,\theta_i+T]$, then, as it was shown in Section 3.2, their incoming vectors-HV are in the relationship corresponding to $\boldsymbol{\psi_{14}}^{-}$. In consequence, during $[\theta_i,\theta_i+T]$ the vector-HV of the whole is given by the projection (filtering) of eq.3.14 into $\psi_{14}^{-}$. From the conditions of orthogonality derived in Section 3.1, the only nonzero term is $\psi_{14}^{-}\otimes\psi_{23}^{-}$. Therefore, systems 2 and 3 show entanglement of the $\psi_{23}^{-}$ type when observed during $[\theta_i, \theta_i+T]$.

In summary: vector-Bell states, which are separable vectors in real space, are able to describe the phenomenon of teleportation thanks to their wave (non-Boolean) nature.

## 4. Kochen-Specker theorem.

### *4.1 Stating the argument.*

Both KS and BI arguments (and also GHZ, see next chapter) demonstrate it is impossible to assign definite values to some quantum observables before the measurements to be carried out are established. But, while the violation of BI is statistical and occurs for a particular state (say, the entangled state $|\varphi_{AB}^{+}\rangle$), KS is a logical impossibility and applies to *any* state (entangled or not). The original KS argument involves a spin-1 system (hence, 3-dimensional) and 117 different orientations in space. An intricate geometrical reasoning demonstrates there is no way to assign values to all observables fitting QM predictions. The essence of the KS argument is much simplified by going up one dimension. Peres [14] and Mermin [15] showed that the same conclusion is reached by considering a system of two spin-½ particles or qubits (hence, 4-dimensional). As my goal here is to provide a description of quantum mysteries as simple as possible, I follow their approach.

| | | |
|---|---|---|
| $\sigma_z^{(1)}$ | $\sigma_z^{(2)}$ | $\sigma_z^{(1)}.\,\sigma_z^{(2)}$ |
| $\sigma_x^{(2)}$ | $\sigma_x^{(1)}$ | $\sigma_x^{(1)}.\,\sigma_x^{(2)}$ |
| $\sigma_z^{(1)}.\,\sigma_x^{(2)}$ | $\sigma_x^{(1)}.\,\sigma_z^{(2)}$ | $\sigma_y^{(1)}.\,\sigma_y^{(2)}$ |

Figure 4: The array of Pauli matrices in the Mermin-Peres version of KS.

Consider then the array of operators' products in Figure 4. The $\sigma_j^{(i)}$ are the Pauli matrices acting on the particle *i*. Regardless the state, the outcome of measuring $\sigma_j^{(i)}$ is always +1 or -1. In each of the rows or columns the observables are mutually commuting, so that they can be measured in any order without perturbing the others or the final result. At first sight, this problem allows a classical description. Let suppose then that it is possible to assign an outcome value (+1 or -1) (before defining the measurements to be performed) to all the $\sigma_j^{(i)}$ in Fig.4. That this assignment (*any* assignment) cannot coincide with QM predictions is easy to demonstrate. Let multiply all columns $C_k$ and rows $R_k$ in Fig.4, then:

$$\Pi_k\, C_k \times R_k = 1 \qquad (4.1)$$

Because each $\sigma_j^{(i)}$ appears twice in the product. According to QM instead, the result of the product in all rows and columns is +1 excepting in the third column, which reads (recall that operators acting on different qubits commute):

$$\sigma_z^{(1)}.\ \sigma_z^{(2)}.\sigma_x^{(1)}.\sigma_x^{(2)}.\sigma_y^{(1)}.\sigma_y^{(2)} = \sigma_z^{(2)}.\ \boldsymbol{\sigma_z^{(1)}.\sigma_x^{(1)}}.\sigma_x^{(2)}.\sigma_y^{(2)}.\sigma_y^{(1)} =$$

$$= \sigma_z^{(2)}.i\sigma_y^{(1)}.i\sigma_z^{(2)}.\sigma_y^{(1)} = i^2\ \sigma_z^{(2)}.\sigma_z^{(2)}.\sigma_y^{(1)}.\sigma_y^{(1)} = -1 \qquad (4.2)$$

Therefore, the supposed assignment of outcomes (no matter which one) cannot coincide with the QM predictions. Note the state upon which measurements are made has not been defined. The impossibility applies to *any* state.

Let see first the cause of that impossibility. The outcome +1 in the two first columns and rows is evident. Let focus in the third row, which reads:

$$\sigma_z^{(1)}.\ \sigma_x^{(2)}.\sigma_x^{(1)}.\sigma_z^{(2)}.\sigma_y^{(1)}.\sigma_y^{(2)} = \sigma_z^{(1)}.\ \boldsymbol{\sigma_x^{(2)}.\sigma_z^{(2)}}.\sigma_x^{(1)}.\sigma_y^{(1)}.\sigma_y^{(2)} =$$

$$= \sigma_z^{(1)}.(-i)\sigma_y^{(2)}.i\sigma_z^{(1)}.\sigma_y^{(2)} = (-i^2)\ \sigma_z^{(1)}.\sigma_z^{(1)}.\sigma_y^{(2)}.\sigma_y^{(1)} = +1 \qquad (4.3)$$

The difference with the result in eq.4.2 (the third column) is because $\sigma_z.\sigma_x = i\sigma_y$, but $\sigma_x.\sigma_z = -i\sigma_y$ (the factors in red and bold type). That is, non-commutation of operators' product. One may wonder if this result can be changed, for all elements in the original array commute with each other. Examination of the arrays' structure shows that this is impossible. In the third column, the $\sigma_z^{(j)}$ (j=1,2) are *both* before or after the corresponding $\sigma_x^{(j)}$ (j=1,2), what always (regardless the order chosen) produces a factor (-1). In the third row instead, if $\sigma_z^{(1)}$ is chosen to be *before* $\sigma_x^{(1)}$ then $\sigma_z^{(2)}$ will appear *after* $\sigma_x^{(2)}$, or vice versa. This difference produces an additional minus sign, and the third row is always equal to +1. Then +1.(-1) = -1.

The fact that Pauli matrices do not commute is not mysterious or "quantum". It is the mere consequence of non-commutation of finite rotations in more than 2 dimensions. The same happens with finite angle rotations in real space in 3-dimensions. In this case too, it does not suffice to say which rotations are going to be applied to predict the outcome. The *order* of the rotations must be established. Be aware that the original KS argument applies for a 3-dimensions space, and the one above for a 4-dimensions space. In a 2-dimensions space rotations do commute and the KS argument does not apply. On the other hand, even in more than 2-dimensions, once the *order* of the rotations is defined, then the final state is well defined.

F.ex, get two cards and write the *x,y,z* axis (*z* going out of the plane of the card upwards). Now rotate one of them $90^o$ with *x* as the rotation axis, in the "positive" direction, that is, by going from *y* to *z*. Then rotate $90^o$ with *z* as the rotation axis, again in the "positive" direction, that is, by going from *x* to *y*. Compare with the card that is not rotated. The rotated *x* points as the original *z*, the rotated *y* points as the original –*x*, and the rotated *z* points as the -*y*. Let start again and change the order of the operations. Apply a $90^o$ rotation with *z* as the axis, going from *x* to *y*, and then a rotation with $x$ as the axis going $90^o$ from $y$ to $z$. The rotated $x$ points now as the original $y$, the rotated $y$ points now as the original $z$ and the rotated $z$ points now as the original $x$. Therefore, by switching the order of the rotations, the final position of the card is completely different. It is impossible to assign numbers (say, the result of projection along some axis) to the final state if the order of the rotations is not previously established. However, saying that this limitation refutes "Realism" is clearly exaggerated. The card and its orientation are always real and well defined. The impossibility of assigning results in advance to arbitrary projections has nothing to do with "quantumness". It is just a consequence of non-commutation of rotations in more than 2 dimensions, a mere geometrical feature.

Pauli matrices correspond to rotations in a 2-dimensional *complex* space, the Poincaré sphere. In order to visualize why a (-1) factor appears when the order of rotations change in eqs.4.2 and 4.3, it is necessary to figure out rotations of complex functions in the Poincaré sphere. This is not easy. In order to digest what essentially happens, I show next an example using rotations in real 3-dimensions space that behaves (almost) exactly as the Pauli matrices in the KS theorem.

*4.2 An example in real 3-Dimensions space.*

One may think that, as $(\sigma_j^{(i)})^2 = 1$, the example in 3-dimensions real space uses rotations in $180^o$. But it does not, because rotations in $180^o$ do commute! Let consider then $90^o$ rotations. The corresponding matrices in standard notation (x,y,z order) are:

$$R_x = \begin{pmatrix} 1 & 0 & 0 \\ 0 & 0 & -1 \\ 0 & 1 & 0 \end{pmatrix}; \quad R_y = \begin{pmatrix} 0 & 0 & 1 \\ 0 & 1 & 0 \\ -1 & 0 & 0 \end{pmatrix}; \quad R_z = \begin{pmatrix} 0 & -1 & 0 \\ 1 & 0 & 0 \\ 0 & 0 & 1 \end{pmatrix} \tag{4.4}$$

In order to mimic the Pauli matrices' feature that their eigenvalues are ±1, let define that a vector ending (after applying an arbitrary rotation) in the $z \leq 0$ $(z>0)$ space corresponds to a -1 (+1) outcome. Now consider the array:

| | | |
|---|---|---|
| $R_z^{(1)}$ | $R_z^{(2)}$ | $R_z^{(1)}. R_z^{(2)}$ |
| $R_x^{(2)}$ | $R_x^{(1)}$ | $R_x^{(1)}. R_x^{(2)}$ |
| $R_z^{(1)}. R_x^{(2)}$ | $R_x^{(1)}. R_z^{(2)}$ | $R_y^{(1)}. R_y^{(2)}$ |

Figure 5: The array of $90^o$ rotations' matrices in 3-dimension real space, identical to Fig.4

This is the same as in Fig.4. The chosen observable is different, though; here it is $(\Pi_k\, C_k \times Row_k)^2$ (see eq.4.1). This is because here the $R_j^{(k)}$ must be raised to the *fourth* power to get the identity. F.ex.: $C_1 = [R_z^{(1)}]^2 \times [R_x^{(2)}]^2 \neq 1$, but $C_1^2 = [R_z^{(1)}]^4 \times [R_x^{(2)}]^4 = 1$.

It seems evident that any assignment of outcomes to the array in Fig.5 should produce a +1

result for the observable $(\Pi_k\ C_k \times Row_k)^2$. In fact, the calculation for all rows and columns is the same as for $C_1$, excepting for the ones with subindex 3. Then, for $C_3$:

$$C_3 = [R_z^{(1)}.\ R_x^{(1)}.R_y^{(1)}].[R_z^{(2)}.\ R_x^{(2)}.R_y^{(2)}];\ \text{then}\ C_3^2 = [R_z^{(1)}.\ R_x^{(1)}.R_y^{(1)}]^2.[R_z^{(2)}.\ R_x^{(2)}.R_y^{(2)}]^2 \quad (4.5)$$

Check that $[R_z^{(k)}.\ R_x^{(k)}.R_y^{(k)}]^2 = 1$, so that $C_3^2 = 1\times 1 = 1$. Instead, $Row_3$:

$$Row_3 = R_z^{(1)}.\ R_x^{(1)}.R_y^{(1)}.\ R_x^{(2)}.\ R_z^{(2)}R_y^{(2)} \Rightarrow Row_3^2 = [R_z^{(1)}.\ R_x^{(1)}.R_y^{(1)}]^2.[R_x^{(2)}.\ R_z^{(2)}R_y^{(2)}]^2 \quad (4.6)$$

The first factor is 1, but the second one is not:

$$[R_x^{(2)}.\ R_z^{(2)}R_y^{(2)}]^2 = \begin{pmatrix} -1 & 0 & 0 \\ 0 & -1 & 0 \\ 0 & 0 & 1 \end{pmatrix} \quad (4.7)$$

which is as close as (-1) as it is possible to go using only rotations. In order to get truly (-1) it is necessary to include an inversion, but I think this example is more elegant if only rotations are used, even at the cost of an imperfect analogy. The matrix in eq.4.7 does invert a vector of the form (a,b,0); for these vectors $(\Pi_k\ C_k \times Row_k)^2 = -1$, and the same impossibility as for the KS argument is demonstrated.

This example shows that also for vectors in real 3-dimensions space it is impossible to assign outcomes (fitting the actual results) to the array in Fig.5 before establishing the order of the rotations. In summary: there is no "quantum" mystery in the KS theorem, just a vectors' feature.

## 5. Greenberger-Horne-Zeilinger states.

### *5.1 Stating the argument.*

The GHZ proof of the incompatibility of QM with classical physics is similar to the KS one. In both cases, assigning set of values fitting all QM predictions is demonstrated impossible before the observations to be carried out are established. However, while the KS case is valid for an arbitrary state, the GHZ case involves a specific state. Following the usual approach [16], consider an entangled state of 3 qubits (in this Chapter, notation of the external product "⊗" is omitted for simplicity):

$$|\Phi^+_{GHZ}\rangle = (1/\surd 2).\{|x_1, x_2, x_3\rangle + |y_1, y_2, y_3\rangle\} \quad (5.1)$$

which is observed at stations 1, 2, 3, see Figure 6. In each station a Pauli spin operator, $\sigma_x$ or $\sigma_y$, is applied. Physically, $\sigma_x$ corresponds to a linear polarizer oriented at $45^o$ of the $\{x,y\}$ axes ($\sigma_l = \sigma_x$), and $\sigma_y$ to a circular polarizer ($\sigma_r = \sigma_y$); then $\sigma_l.|x\rangle = |y\rangle$, $\sigma_l.|y\rangle = |x\rangle$, $\sigma_r.|x\rangle = i|y\rangle$, $\sigma_r.|y\rangle = -i|x\rangle$. The operator to be applied is arbitrarily chosen in each station. The set of operators applied to $|\Phi^+_{GHZ}\rangle$ is named a *configuration*. F.ex., $\sigma_r^{(1)}.\sigma_r^{(2)}.\sigma_l^{(3} \equiv rrl$ is a possible configuration. It means that $\sigma_r$ is applied in stations 1 and 2 and $\sigma_l$ in station 3 (as in Fig.6).

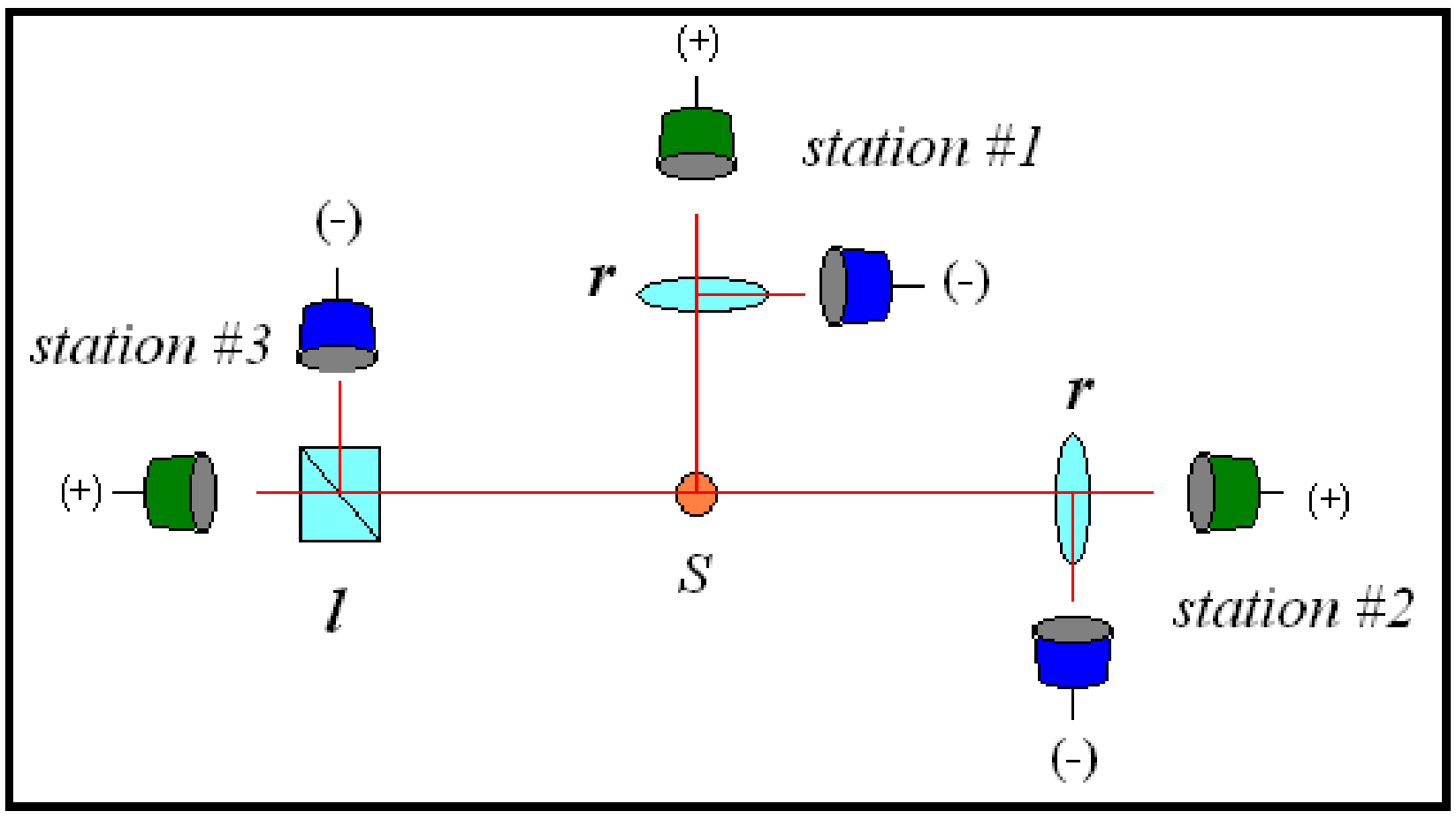

Figure 6: Scheme of a setup to observe $\sigma_r^{(1)}.\sigma_r^{(2)}.\sigma_l^{(3)}$ when the state is $|\Phi^+_{GHZ}\rangle$ (eq.5.1). An ellipse represents a circular polarization analyzer (operator $\sigma_r$, or $\sigma_y$). A square represents a linear one set at 45° (operator $\sigma_l$, or $\sigma_x$). For the displayed operators' configuration (*rrl*) the product of the outcomes of the three stations is always (-1). Yet, the outcomes in each station are undefined; they correspond to non-polarized beams. The correlation among the results (stations can be far from each other) cannot be explained by a set of classical instructions.

In each station, a balanced number of (+1) or (-1) outcomes is predicted. But, the expected value of the *product* of the three outcomes, f.ex:

$$\langle\Phi^+_{GHZ}|\sigma_r^{(1)}.\sigma_r^{(2)}.\sigma_l^{(3)}|\Phi^+_{GHZ}\rangle = \langle\Phi^+_{GHZ}|\{i^2.|y_1, y_2,y_3\rangle + (-i)^2.|x_1, x_2,x_3\rangle\}/\sqrt{2} = -1 \quad (5.2)$$

This result is predicted to occur with certainty. This is because $|\Phi^+_{GHZ}\rangle$ is an eigenstate of the operators' product with the *rrl* configuration, with eigenvalue (-1). The same happens with *rlr* and *lrr*; the configuration *lll* has eigenvalue (+1) instead. For the remaining four configurations (f.ex.: *lrl*), $|\Phi^+_{GHZ}\rangle$ is not an eigenstate, and the total result is (+1) or (-1) with equal probability. Actually, in these cases the state obtained after applying the corresponding operators is orthogonal to $|\Phi^+_{GHZ}\rangle$, so that the expected value of the total result is zero. It is impossible to get zero by multiplying ±1 outcomes, hence this zero value is the average obtained after a statistically significant number of observations.

A set of instructions fitting all QM predictions for $|\Phi^+_{GHZ}\rangle$ is impossible. Consider a classical model that uses 2×3 matrices to assign the outcomes in each station depending on the operator that is applied, say:

| Analyzer found | Outcome in station 1 | Outcome in station 2 | Outcome in station 3 |
|---|---|---|---|
| *l* | $l_1 = +1$ | $l_2 = +1$ | $l_3 = +1$ |
| *r* | $r_1 = +1$ | $r_2 = -1$ | $r_3 = -1$ |

Figure 7: Matrix table of instructions (or hidden variables) for a GHZ state.

This matrix is able to reproduce QM predictions for configurations *lll*, *rrl*, and *rlr*, but it fails with *lrr*. It cannot be done better: using reasoning similar to the one in the KS theorem, the product of the set of instructions for the four configurations stated above (the ones for which $|\Phi^{+}_{GHZ}\rangle$ is eigenstate) is: $(l_1l_2l_3).(r_1r_2l_3).(r_1l_2r_3).(l_1r_2r_3) = (l_1.l_2.l_3.r_1.r_2.r_3)^2 = 1$, but the product of the four corresponding QM predictions is always $(-1)^3\times(+1) = (-1)$. Therefore, no set of instructions can reproduce all QM predictions. This demonstrates the logical incompatibility between classical physics and QM for the state $|\Phi^{+}_{GHZ}\rangle$, but it *does not* provide a "single shot" refutation of classical physics, as it is often claimed. For, each matrix like the one in Fig.7 can reproduce the QM predictions for 7 of the 8 possible configurations. Assuming that configurations applied by the observer and matrices carried by the particles are uncorrelated (that is, assuming "measurement independence") the probability of observing a discrepancy from QM predictions is 1/8 per trio in the ideal case. In consequence, refutation of classical physics can be only statistical, as in the BI's case. In the ideal case GHZ states can provide, instead, a "single shot" refutation *of QM* [17].

By the way, the GHZ state of 2 qubits is the Bell state $|\varphi^{+}_{AB}\rangle$ discussed in Chapter 2. The configuration *ll* has total product (+1) and *rr* (-1), while *lr* and *rl* produce total results that average to zero. Yet, for this case, 2×2 matrices of instructions similar to Fig.7 reproducing QM predictions for all the configurations are possible, and easy to find.

*5.2 Description in terms of vector-HV.*

The KS proof is based on the non-commutation of the product of Pauli operators (ultimately, on the non-commutation of rotations in more than 2 dimensions). Here, instead, the operators act on different qubits. Therefore, all of them commute; they can be applied in any order. And it must be so, because the three stations can be at arbitrary distances of each other and the source (i.e., detection events can be space-like separated). It is therefore impossible defining an absolute order (i.e., valid for all systems of reference) of each observation with respect to the others. Nevertheless, the GHZ theorem is not as exotic as it may appear. It is just vectors' geometry in a high dimensional ($2^3 = 8$) *product* space. Visualizing what happens in this realm is impossible, but it is possible to see that QM predictions for $|\Phi^{+}_{GHZ}\rangle$ have nothing specifically "quantum". Consider a classical vector-HV in the product space of three two-dimensional vectors (the same notation than in Chapter 2):

$$\mathbf{V_{GHZ}}(t) \equiv V(t).(\mathbf{e}_{x1}.\mathbf{e}_{x2}.\mathbf{e}_{x3} + \mathbf{e}_{y1}.\mathbf{e}_{y2}.\mathbf{e}_{y3}) /\sqrt{2} \quad (5.3)$$

It behaves exactly as $|\Phi^{+}_{GHZ}\rangle$, f.ex.:

$$\sigma_r^{(1)}.\sigma_r^{(2)}.\sigma_l^{(3)}.\mathbf{V_{GHZ}}(t) = V(t).\{i^2.\mathbf{e}_{y1}.\mathbf{e}_{y2}.\mathbf{e}_{y3} + (-i)^2.\mathbf{e}_{x1}.\mathbf{e}_{x2}.\mathbf{e}_{x3}\} = -\mathbf{V_{GHZ}}(t) \quad (5.4)$$

and so for all the other configurations.

Therefore, the core of the GHZ theorem is that a rotated vector (here, a vector in a 8-dimensions space) may end parallel, or anti-parallel, or perpendicular to the non-rotated vector depending on the *number and type* of rotations applied. This is hardly surprising! Consider the following simple example: a vector **v** in the real *plane*, and rotations $R_\pi$ (rotation in $\pi$) and $R_{\pi/2}$. Now replace Pauli operators $\{\sigma_r, \sigma_l\}$ applied to qubits #1, #2, #3 of $|\Phi^+_{GHZ}\rangle$, with $\{R_{\pi/2}, R_\pi\}$ applied to **v** in order #1, #2, #3; in both cases all operations commute. Note that:

$$(R_\pi.R_\pi.R_\pi.\mathbf{v}).\mathbf{v} = -1 \tag{5.5a}$$

$$(R_\pi.R_\pi.R_{\pi/2}.\mathbf{v}).\mathbf{v} = 0 \tag{5.5b}$$

$$(R_\pi.R_{\pi/2}.R_{\pi/2}.\mathbf{v}).\mathbf{v} = 1 \tag{5.5c}$$

which are the same results as with $|\Phi^+_{GHZ}\rangle$ and which, as before, cannot be fitted by a set of previously defined instructions. The *number and type* of rotations of each class determines the result. In this sense, the core of the GHZ theorem is trivial. Nature of Pauli operators, which correspond to rotations in a complex sphere, and the 8-dimensions of the problem are the reasons that make it difficult to visualize.

The QM predictions for the number of coincidences can be fitted by projecting $\mathbf{V_{GHZ}}(t)$ on the corresponding axes, and integrating the modulus squared of the projected component in time, as it was done in Chapter 2. F.ex., let calculate the number of triple detections with outcomes (-1) in station #1 and station #2, and (+1) in station #3 when the configuration is *llr*:

$$N_{llr}^{--+} = (1/u). \int_0^{Tr} dt.\, |\mathbf{e}_{-1}.\mathbf{e}_{-2}.\mathbf{e}_{R3}.\mathbf{V_{GHZ}}(t)|^2 \tag{5.6}$$

where $\mathbf{e}_{-1}$ corresponds to anti-diagonal linear polarization, because the outcome which frequency we want to calculate in this station is (-1) and the polarizer in station #1 is oriented at 45°, the same with $\mathbf{e}_{-2}$, and $\mathbf{e}_{R3}$ corresponds to circular right-hand polarization, because the polarizer in station #3 is circular and the outcome we want to calculate in this station is (+1).

It is convenient writing $\mathbf{V_{GHZ}}(t)$ in the diagonal-polarized basis:

$$\mathbf{V_{GHZ}}(t) = ½.V(t).\{\mathbf{e}_{+1}.\mathbf{e}_{+2}.\mathbf{e}_{+3} + \mathbf{e}_{+1}.\mathbf{e}_{-2}.\mathbf{e}_{-3} + \mathbf{e}_{+1}.\mathbf{e}_{-2}.\mathbf{e}_{+3} + \mathbf{e}_{-1}.\mathbf{e}_{-2}.\mathbf{e}_{+3}\} \tag{5.7a}$$

and in the circular-polarization basis:

$$\mathbf{V_{GHZ}}(t) = ¼.V(t).\{(1-i).[\mathbf{e}_{L1}.\mathbf{e}_{L2}.\mathbf{e}_{L3} + \mathbf{e}_{L1}.\mathbf{e}_{R2}.\mathbf{e}_{R3} + \mathbf{e}_{R1}.\mathbf{e}_{L2}.\mathbf{e}_{R3} + \mathbf{e}_{R1}.\mathbf{e}_{R2}.\mathbf{e}_{L3}] + \\ + (1+i).[\mathbf{e}_{R1}.\mathbf{e}_{R2}.\mathbf{e}_{R3} + \mathbf{e}_{R1}.\mathbf{e}_{L2}.\mathbf{e}_{L3} + \mathbf{e}_{L1}.\mathbf{e}_{R2}.\mathbf{e}_{L3} + \mathbf{e}_{L1}.\mathbf{e}_{L2}.\mathbf{e}_{R3}]\} \tag{5.7b}$$

note corresponding expressions apply also for $|\Phi^+_{GHZ}\rangle$ and bases' states in ket notation.

If calculation is carried out in station #1 (of course, the number of coincidences is the same in all stations), then applying $\mathbf{e}_{R3}$ to $\mathbf{V_{GHZ}}(t)$ in eq.5.7b, and then $\mathbf{e}_{-2}$ to the result, one gets $½.\mathbf{e}_{R1}.V(t)$. Helpful expressions are displayed in Section A.3 in the Appendix. Then:

$$N_{llr}^{--+} = (1/u). \int_0^{Tr} dt.\, |\mathbf{e}_{-1}.½.\mathbf{e}_{R1}.V(t)|^2 = |(1-i)/4|^2.\, (1/u). \int_0^{Tr} dt.\, V^2(t) = N/8 \tag{5.8}$$

which is the correct result, because for *llr* the total product has no definite sign, thus coincidences are shared equally among the 8 possible combinations of ±1 outcomes. The same applies to all combinations of outcomes for configurations *rll*, *lrl* and *rrr*. For *rlr*, instead (find the expression of $\mathbf{e}_{-2}.\mathbf{e}_{R3}.\mathbf{V_{GHZ}}(t)$ in the Appendix eq.A.3.8):

$$N_{rlr}^{--+} = (1/u).\int_0^{Tr} dt.\,|\mathbf{e}_{L1}.\mathbf{e}_{-2}.\mathbf{e}_{R3}.\mathbf{V_{GHZ}}(t)|^2 = (1/u).\int_0^{Tr} dt.\,|\mathbf{e}_{L1}.½.\mathbf{e}_{R1}.V(t)|^2 = 0 \qquad (5.9)$$

(because $\mathbf{e}_{L1}.\mathbf{e}_{R1} = 0$), which also fits the QM prediction, and the same result is obtained for $N_{rlr}^{-+-}$, $N_{rlr}^{+--}$, $N_{rlr}^{+++}$. For $N_{rlr}^{-++}$ instead:

$$(1/u).\int_0^{Tr} dt.\,|\mathbf{e}_{L1}.\mathbf{e}_{+2}.\mathbf{e}_{R3}.\mathbf{V_{GHZ}}(t)|^2 = (1/u).\int_0^{Tr} dt.\,|\mathbf{e}_{L1}.½.\mathbf{e}_{L1}.V(t)|^2 = ¼.(1/u).\int_0^{Tr} dt.\,V^2(t) = N/4 \qquad (5.10)$$

(outcome -1 in #1 means projection on $\mathbf{e}_L$ because the polarizer in #1 is *r*, outcome +1 in #2 means projection on $\mathbf{e}_+$ because the polarizer in #2 is *l*, outcome in #3 means projection on $\mathbf{e}_R$ because the polarizer in #3 is *r*). The same result is obtained for $N_{rlr}^{+-+}$, $N_{rlr}^{++-}$ and $N_{rlr}^{---}$, so that the $N_{rlr}^{ijk}$ ($i,j,k = \pm1$) sum up $4\times N/4 + 0\times N/4 = N$, as expected. The same applies to all the $N_{rrl}^{ijk}$ and $N_{lrr}^{ijk}$.

For the $N_{lll}^{ijk}$ the values are swapped: $N_{lll}^{-+-}$, $N_{lll}^{+--}$, $N_{lll}^{+++}$, $N_{lll}^{--+} = N/4$, and $N_{lll}^{-++}$, $N_{lll}^{+-+}$, $N_{lll}^{++-}$ and $N_{lll}^{---} = 0$. Therefore, observable number of triple coincidences of each type and for each configuration are obtained by merely projecting $\mathbf{V_{GHZ}}(t)$ on the corresponding axes.

The GHZ theorem is often regarded as the ultimate proof of “quantum non-Locality”. Without non-Locality, it is argued, how do the detectors in each station in Fig.6 know whether they have to fire, or not, in order to fit the QM prediction *for each trio,* if they ignore both the configuration and what has happened at the other stations? Nevertheless, note that explaining what happens *for each trio* corresponds to the “hard” problem mentioned in Chapter 2. The hard problem is presumably solvable with a new type of quantum computer only. The “soft” problem, that is, obtaining correct *statistical* predictions, is solved simply by projecting the classical $\mathbf{V_{GHZ}}(t)$ on the corresponding axes, as it is just demonstrated (also QM solves the soft problem, of course).

### *5.3 Vectors in product space.*

The (statistical) results derived in the previous Section do not involve any kind of non-locality. They are obtained by merely projecting an 8-dimensional vector on different axes, as it was done in Chapter 2 with the vector-Bell states. The results depend on the outcomes to be observed, what determines the axes of projection, and on the rotations that have been previously applied to said vector. This is hardly surprising, and there is nothing “quantum” here, just 8-dimensions geometry. Note, however, an important difference with the calculations in Chapter 2: in that Chapter, the BI are violated by two related but separate classical vectors that exist in normal space, $\mathbf{V^A}(t)$ and $\mathbf{V^B}(t)$. The total problem is described in a space that is the *sum* of two 2-dimensions spaces. Here, instead, $\mathbf{V_{GHZ}}(t)$ is a vector in a *product* space. True, $|\varphi^+_{AB}\rangle$ is also a vector in a product space, but in the previous Chapters it is demonstrated unnecessary to explain the violation of BI, quantum teleportation, etc. The important question now is whether a vector (regardless if it is of quantum nature as $|\Phi^{+}_{GHZ}\rangle$, or classical nature as $\mathbf{V_{GHZ}}$) belonging to a *product* space can actually exist [18]. One way to show that it can indeed exist (at least, that something that behaves like it, can exist) is to follow, step by step, the procedure followed in the experimental verification of the GHZ theorem [19].

The setup is sketched in Figure 8. We are interested only in the cases *when the four detectors* (T, D1, D2, D3) *fire simultaneously*. Radiation from a pulsed UV-laser produces, in a single nonlinear crystal, two near-IR photons with orthogonal polarizations that propagate along diverging paths *a* and *b*. Four simultaneous detections necessarily means that the pump pulse has produced two (non-entangled) pairs of near-IR photons, say: ***Ha*** and ***Va***, and ***Hb*** and ***Vb***. They are usually represented by kets, but they can be interpreted as vector-HV as well (notation of vector-HV time dependence is omitted in what follows).

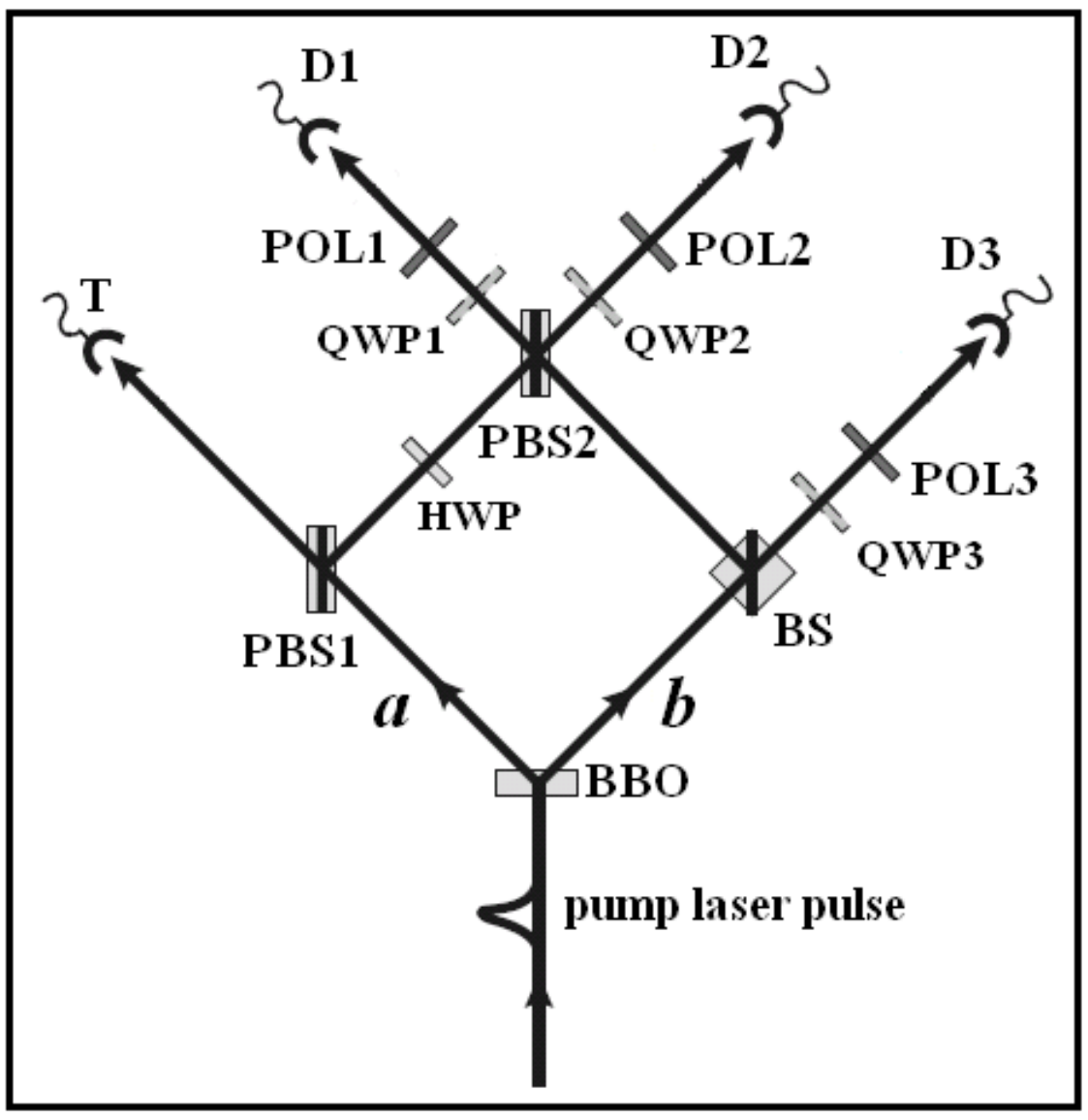

Figure 8: (adapted from [19]). Scheme of the experimental set-up that generates $|\Phi^{+}_{GHZ}\rangle$. This occurs only when the four single photon detectors T, D1, D2, D3 fire simultaneously, i.e., generation is based on post-selection of data. BBO: nonlinear crystal, PBS: polarization beam splitters, BS: beam splitter, POL: polarizers, HWP: half wavelength waveplate, QWP: quarter wavelength waveplates.

Let suppose that a photon is detected at station T ("trigger"). This photon has been transmitted through polarizer PBS1, so it must be a horizontally polarized one, let's call it ***Ha***. In consequence, its "brother" ***Va*** has followed the other path (*b*, rightwards), towards beam-splitter BS. Therefore, if a photon is reflected at polarizer PBS1, it must be the vertical photon of the other pair, ***Vb*** (if no photon is reflected at PBS1, then it is impossible the four detectors to fire simultaneously, and detections produced by this pump pulse are not taken into account). In consequence, ***Hb*** has followed path *b* towards BS. Photon ***Vb*** is rotated by half wavelength waveplate HWP to a state of diagonal polarization: $(1/\sqrt{2})(H'+V')$. We assume that detectors D1 and D2 fire simultaneously. This is possible if:

1) The component ***V'*** of ***Vb*** is reflected by PBS2 towards D1, and ***Va*** (which had followed the other path and was reflected at the BS), is reflected by PBS2 towards D2.

OR:

2) The component ***H'*** of ***Vb*** is transmitted by PBS2 to D2, and ***Hb*** (which had followed the other path and was reflected at the BS), is transmitted by PBS2 to D1.

In the case 1), if D3 also fires, it occurs because ***Hb*** was transmitted at the BS and went to D3.
In the case 2), if D3 also fires, it occurs because ***Va*** reached the detector instead.

Of course there are other possibilities, but in none of them the four detectors fire simultaneously. Summing up the two possibilities above, the total vector-HV amplitude is: (1/√2).***V'.Va.Hb*** + (1/√2).***H'.Hb.Va***. It only remains defining the axes of polarization at D3 rotated 90$^o$ with respect to the others (by adjusting QWP3), and $\mathbf{V_{GHZ}}$ (or $|\Phi^+_{GHZ}\rangle$) is obtained in the paths 1 to 3. Afterwards, orientations of POL1…POL3 and QWP1…QWP3 are adjusted to check that the obtained state does behave as $|\Phi^+_{GHZ}\rangle$ or $\mathbf{V_{GHZ}}$.

In the previous lines the usual ket notation is replaced by the vector-HV one (omitting the time dependence). This is to make evident that the above discussion applies even if each amplitude is defined to be a classical time varying vector **V**(t) as in Chapter 2. The only difficulty is in polarizer P2, when "half" the vector ***Vb*** is transmitted and the other "half" is reflected. In this case the additional instruction "winner-take-all" of Section 3.2 (i.e., the "full" photon follows the path with the vector-HV having the highest amplitude) solves the difficulty. If the two amplitudes are *exactly* equal, D1 or D2 do not fire, and that pump pulse is simply not taken into account.

In summary: the apparently paradoxical features of GHZ states arise from geometry in a high dimensional space. There is nothing specifically "quantum"; the classical HV-vector $\mathbf{V_{GHZ}}$(t) behaves in the same way than $|\Phi^+_{GHZ}\rangle$ and produces the same observable results. The intriguing question is whether vectors in product space can actually exist. Something that behaves like them can be built from a particular superposition of classical vectors plus post-selection.

**Conclusions.**

Violation of the BI is described, without violation of Locality or Realism, in terms of separable classical vectors in real space. Of course, *non-Boolean versions* of Locality and Realism, which avoid using probabilities in their definition [2]. The key is that "filtering" in the proposed non-Boolean algebra means projection of vectors, instead of intersection of sets. These vectors (here called vector-Bell states) can also describe Hong-Ou-Mandel effect and Teleportation. These effects are the consequence of *superposition, interference, orthogonality* and *projection*, which are classical vectors' features.

The KS theorem is the consequence of *non-commutation of rotations* in more than two dimensions, a well known vectors' feature. An example with rotations in $\pi/2$ in three dimensions behaves (almost) exactly as the KS theorem. In both cases, it is impossible assigning outcomes to measurements before the order of the rotations to be applied is established. Note that rotation of a Boolean HV (as $\lambda$ in Section 2.2) is meaningless.

The features of GHZ states can be understood as the consequence that a rotated vector may end parallel or perpendicular to the original vector, depending on how many rotations of each type are applied, what is hardly surprising. All statistical predictions of QM can be explained as the consequences of *projections* on different axes of a vector in a high dimensional product space. The question that remains is whether vectors in a product space exist in the real world or not. It is shown that the behavior expected for such vectors is reproduced by superposition of vectors in normal space plus post-selection of data.

I hope the Reader finds the discussed quantum mysteries more digestible now.

Let comment on the origin of quantum mysteries. Some 25 centuries ago, Democritus proposed the physical world to be made of a huge number of point-like particles in permanent motion. Observed changes would be the consequence of the different ways they combine or aggregate. They would evolve according to some rules or blind processes which, in his time, had yet to be elucidated. Particles can be classified and gathered in sets (thus, they follow intuitive Boolean logic). Measures can be defined over these sets, and then probabilities of finding particles inside some sets, and probability distributions according to the information available on the particles' features.

Democritus' hypothesis was extremely powerful, but it turned out to be incomplete. It cannot explain the phenomenon of interference, which he never knew. Waves are the natural way to understand interference. It is worth remembering here that QM is built as it is, because it has to describe wave phenomena. The condition of stationary waves explains the lack of radiated energy from accelerated electrons in the atom and the discrete pattern of spectral lines. Interference of elementary particles is directly observed. Schrödinger equation is a wave equation. Microphysical phenomena are undoubtedly *wavy*.

Vectors are the simplest mathematical tool to describe waves. Vectors' algebra is non-Boolean, and hence can violate Boole's conditions of completeness (in this case, the BI). Rotations of vectors in more than 2 dimensions do not commute, so that one has to establish the order of the rotations in order to calculate the vector's final orientation and the outcomes' values (KS theorem). Vectors in product space can produce apparently "nonlocal" effects when observed from the normal, separable space. These are not *quantum* mysteries; they are just the consequence of geometry and vectors' features. One should not speak of non-Locality or non-Realism of QM, but of "*vector-iness*" or, to be more general, of "non-Booleanity" of QM.

These soothing arguments do not mean that the common perception of the existence of mysteries in QM is misled. For, vectors are strange things, stranger than they seem to be at first sight. Several quantum mysteries are just the ones of vectors, but vectors are, for our Boolean intuition, mysterious enough. The reasons of some of these perplexities are commented in the Appendix next.

**APPENDIX.**

*A.1 Some commentaries on vectors' mysteries: Filtering vectors.*

Replacing sets by vectors is more disruptive than it may seem. It immediately leads to anti-intuitive results. Consider, f.ex., a system that, when examined, is made of spheres (S) and cubes (C) that can be blue (B) or red (R). No sphere is a cube, and no red object is blue, but there are blue spheres and red spheres; the same happens with the cubes. If Boolean algebra can be applied, then the system is represented by sets, see Figure 6. The measure of the intersections depends on the rate of spheres and cubes that are found to be blue or red.

If non-Boolean vectors' algebra is to be applied instead, the system is represented by vectors along two pairs of orthogonal axes {**B**,**R**} and {**S**,**C**}. The axes {**S**,**C**} are oriented at an intermediate angle with the {**B**,**R**} ones. This angle's value depends on the rate of spheres and cubes that are found to be blue or red.

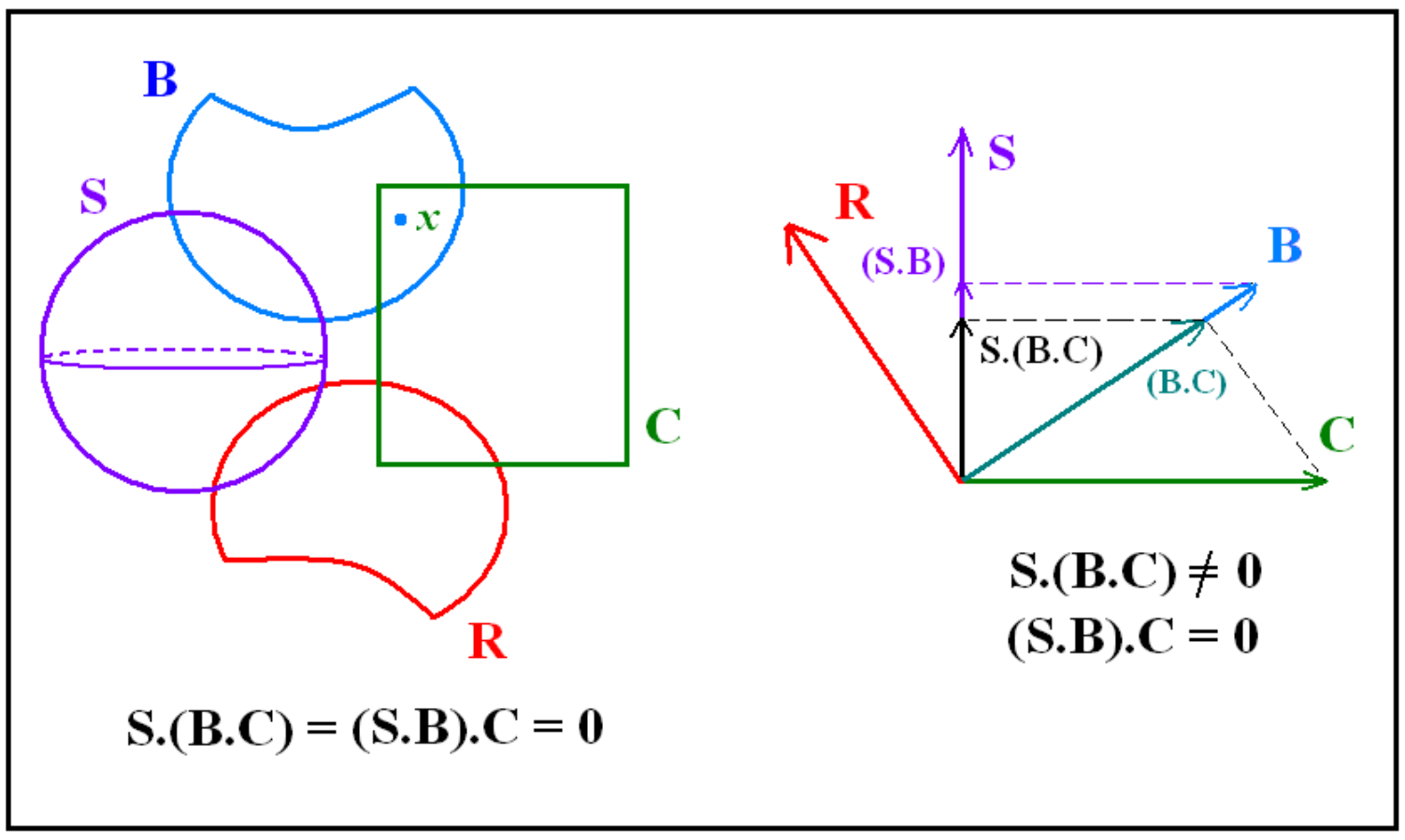

Figure A.1: Red and blue cubes and spheres, according to Boolean and vectors' (non-Boolean) logics.

In the Boolean framework, if filtering (intersection) is performed, no red object is blue $B \cap R=0$ and no sphere is a cube $S \cap C=0$, regardless the order of filtering chosen. In the non-Boolean vectors' framework instead, filtering means projection. On the one hand, **S.C**= 0 as before. But, it may also be: **S.(B.C)** $\neq$ 0. This means that one starts with red and blue cubes and, after filtering out the red objects, ends with red and blue *spheres*.

Elements in the set C (as the element "*x*") do not lose their individuality when the intersection $B \cap C$ is performed. All elements can always be recognized. In the vectors' realm, instead, it is like having a "liquid". When the projection of **C** into **B** is made, all what is projected has identical features, fully described by **B**. Individuality is lost. Things are not cubes or spheres (or any other object) any longer, they are just *blue*. If a new projection is performed, cubes and spheres (or any other object) can be retrieved, but this cannot be known *in advance*. There is no quantum mystery here, just a vectors' feature. This feature is present in the violation of the BI described in Chapter 2: once the vector **V**(t) is filtered along the axis $\mathbf{e}_\alpha$, the filtered vector (*all* of it) is parallel to $\mathbf{e}_\alpha$, and further filtering along the axis $\mathbf{e}_\beta$ produces the factor $cos(\alpha\text{-}\beta)$, which enters the region forbidden by Boolean logic (the BI) in Fig.3.

*A.2 Some commentaries on vectors' mysteries: Contextuality.*

"Non-Booleanity" does not imply that a description in terms of classical probabilities is always impossible. It just means that such a description cannot be taken for granted. The violation of BI is one example of the paradoxes that arise when Boolean algebra is applied when it is impossible. Vorob'yev's theorem [20] demonstrates that the description of non-Boolean problems with Boolean tools can be safely achieved only by using *contextual* scenarios. A few words on the meaning of contextuality are in order.

Let assume that the space Ω of the HVs is assumed a classical probability space. Averages over it allow calculating observable probabilities. Contextuality then means that it is necessary to define a Boolean sub-algebra $\Omega_B$ for each physical quantity to be observed [21]. These sub-algebras define the different contexts or, in the practice of Bell experiments, the angle settings (α,β in Fig.1).

The origin of contextuality can be visualized as follows. One may think that the angle variable ν(t) of the non-polarized vector **V**(t) in Chapter 2 has a uniform distribution, see Figure A.2a. Yet, as a consequence of the principle of superposition (a vectors' feature) the distribution of ν(t) can also be thought as two (half) Dirac's deltas centered at π/2 of each other, see Fig.A.2b. Which one of these distributions is the correct one? The uniform one or the two deltas? (and in the latter case, which one of the infinite possible pairs?). In the Boolean way of thinking, only an infinitesimal fraction of the uniformly distributed values of ν(t) is parallel or orthogonal to a certain angle value α. Yet, this fraction turns out to be here as large as ½, and for *any* value of α. How did the photon know which setting I was going to choose? This situation has the flavor of an inversion of cause and effect, a perplexity often found in QM.

But there is no such inversion. The correct answer is: no probability distribution can be defined. The definition of (classical) probability presupposes Boolean algebra, and vectors follow a non-Boolean one. A probability distribution cannot be properly assigned because of the non-Boolean superposition principle, i.e., the possibility to project **V**(t) into arbitrary axes. Yet, once the value of α is defined (i.e., once the context is defined), probabilities can be assigned. Besides, *once the context is defined* and applying a threshold condition, it is possible to know whether a particle will be detected at a certain gate at a certain time, or not, even *before* **V**(t) reaches the polarizer, see Fig.A.2c. It is possible to know even if a particle *would be* detected or not, if a polarizer were in place. The definition of the context allows applying familiar Boolean logic.

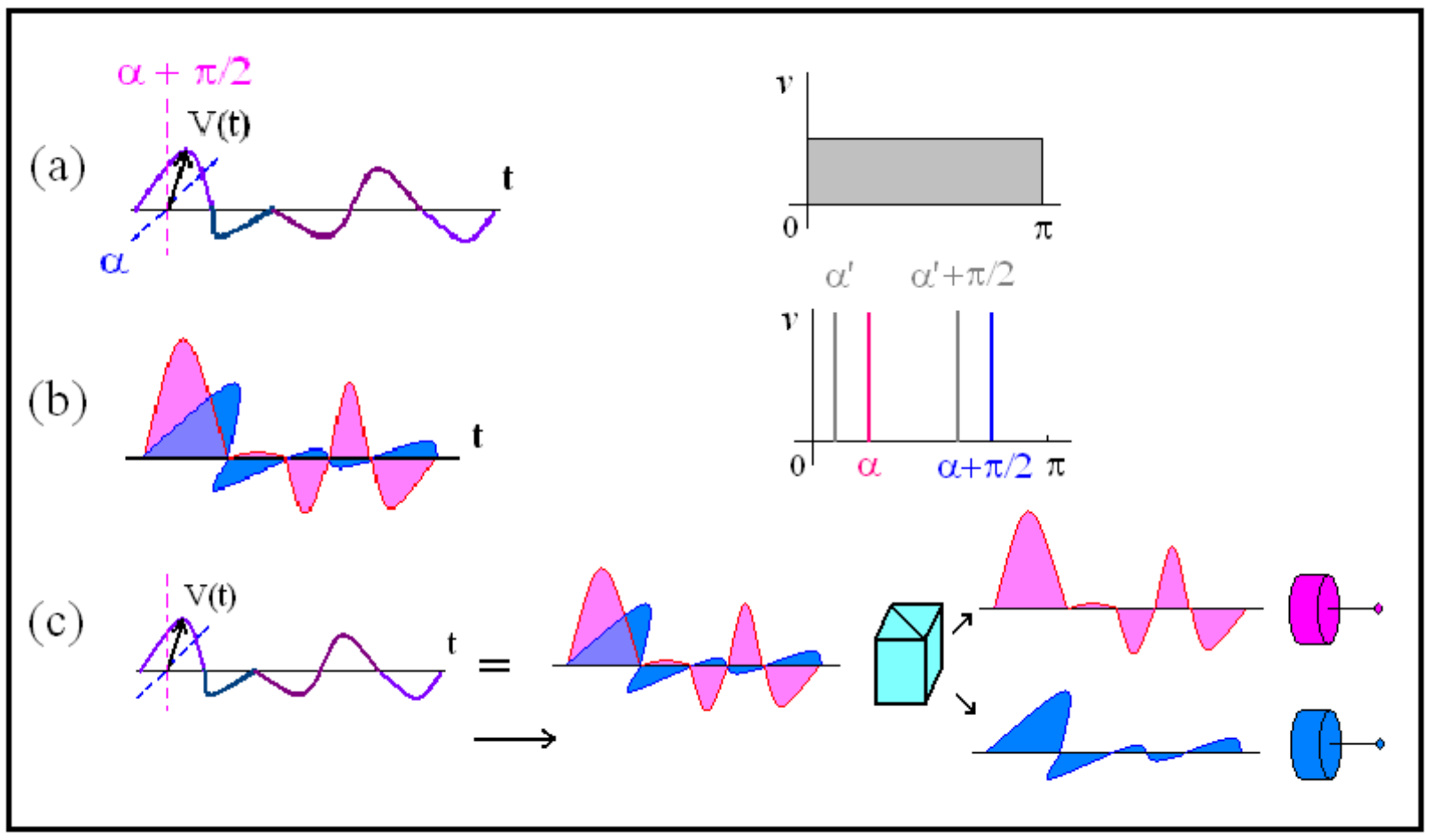

Figure A.2: **(a)** *Left:* vector **V**(t) evolves inside and outside the plane of the paper (red in the plane, blue orthogonal to it, different degrees of violet mean intermediate planes). *Right:* For a non-polarized beam, the "probability distribution" of the vector's angle ν(t) is intuitively thought uniform in [0,π]. **(b)**: *Left:* **V**(t) can be projected into two components, parallel and orthogonal to an arbitrary angle α. *Right:* as a consequence, the "probability distribution" of ν(t) changes from uniform to two "deltas", placed at α and α+π/2. Yet, α' and α'+π/2 are also possible. This does not mean an inversion of cause and effect, but that a description in terms of probability distributions (which presupposes Boolean logic) is impossible. **(c)**: Once the value of α (i.e., the context) is defined, the vector components that determine whether one detection in each output gate of an analyzer occurs, or not, are well defined even before reaching the analyzer.

This example hopefully makes digestible the often stated quantum principle: *the observed properties of a physical system depend on the definition of the apparatus of observation* (in Fig.A.2, this means defining the value of α). This is evident in the vector's realm, and it does not imply violation of Realism: **V**(t) is always well (counterfactually) defined. But, if photon detection after a polarizer is attempted to be described with Boolean logic, then the value of α, i.e. the context (the $\Omega_B$ sub-algebra) must be defined. There is no such thing as a metaphysical influence of the presence of an observer, just the necessity of defining the axes. Be aware that the definition of the context does not mean a physical interaction. The ideal polarizer in Fig.A.2 does not physically interact with **V**(t) (no exchange of energy or momentum occurs). It merely splits two already existing vector's components.

A simple way to retrieve a Boolean framework to deal with non-Boolean situations is to accept *extended* probabilities, i.e., probability values outside the Kolmogorovian (or classical) interval [0,1] [22]. This alternative is well known in Quantum Optics, where a Wigner distribution function taking negative values is a customary indication of non-classical features. Not surprisingly, extended probabilities are related with contextuality. The difference between the total sum of the probability's modulus and 1 (which is, of course, the total sum in classical probability) defines a *coefficient of contextuality* [23,24]. Other alternative "apparently Boolean" descriptions of Bell's experiment also involve not-Kolmogorovian probabilities, such as probabilities defined over singular measures [25] or over non-measurable sets [26], or p-adic ultrametric distances [27-29].

*A.3 Useful expressions to project* $\mathbf{V}_{\mathbf{GHZ}}(t)$**.**

From eq. 5.7a:

$$\mathbf{e}_{+3}.\mathbf{V}_{\mathbf{GHZ}}(t) = \tfrac{1}{2}.\{\mathbf{e}_{+1}.\mathbf{e}_{+2} + \mathbf{e}_{-1}.\mathbf{e}_{-2}\}.V(t) \qquad (A.3.1)$$

$$\mathbf{e}_{-3}.\mathbf{V}_{\mathbf{GHZ}}(t) = \tfrac{1}{2}.\{\mathbf{e}_{+1}.\mathbf{e}_{-2} + \mathbf{e}_{-1}.\mathbf{e}_{+2}\}.V(t) \qquad (A.3.2)$$

and corresponding expressions if the first projection is performed on stations #1 or #2. The symmetry of these expressions (a consequence of the GHZ symmetry) implies that projections on the different particles' spaces, in this case, do commute.

From eq.5.7.b:

$$\mathbf{e}_{R3}.\mathbf{V}_{\mathbf{GHZ}}(t) = \tfrac{1}{4}.\{(1-i).[\mathbf{e}_{L1}.\mathbf{e}_{R2} + \mathbf{e}_{R1}.\mathbf{e}_{L2}] + (1+i).[\mathbf{e}_{R1}.\mathbf{e}_{R2} + \mathbf{e}_{L1}.\mathbf{e}_{L2}]\}.V(t) \qquad (A.3.3)$$

$$\mathbf{e}_{L3}.\mathbf{V}_{\mathbf{GHZ}}(t) = \tfrac{1}{4}.\{(1-i).[\mathbf{e}_{L1}.\mathbf{e}_{L2} + \mathbf{e}_{R1}.\mathbf{e}_{R2}] + (1+i).[\mathbf{e}_{R1}.\mathbf{e}_{L2} + \mathbf{e}_{L1}.\mathbf{e}_{R2}]\}.V(t) \qquad (A.3.4)$$

The results of the next projection then read:

$$\mathbf{e}_{+2}.\mathbf{e}_{+3}.\mathbf{V}_{\mathbf{GHZ}}(t) = \tfrac{1}{2}.\mathbf{e}_{+1}.V(t) = \mathbf{e}_{-2}.\mathbf{e}_{-3}.\mathbf{V}_{\mathbf{GHZ}}(t) \qquad (A.3.5)$$

$$\mathbf{e}_{-2}.\mathbf{e}_{+3}.\mathbf{V}_{\mathbf{GHZ}}(t) = \tfrac{1}{2}.\mathbf{e}_{-1}.V(t) = \mathbf{e}_{+2}.\mathbf{e}_{-3}.\mathbf{V}_{\mathbf{GHZ}}(t) \qquad (A.3.6)$$

$$\mathbf{e}_{+2}.\mathbf{e}_{R3}.\mathbf{V}_{\mathbf{GHZ}}(t) = \tfrac{1}{2}.\mathbf{e}_{L1}.V(t) = \mathbf{e}_{-2}.\mathbf{e}_{L3}.\mathbf{V}_{\mathbf{GHZ}}(t) \qquad (A.3.7)$$

$$\mathbf{e}_{+2}.\mathbf{e}_{L3}.\mathbf{V}_{\mathbf{GHZ}}(t) = \tfrac{1}{2}.\mathbf{e}_{R1}.V(t) = \mathbf{e}_{-2}.\mathbf{e}_{R3}.\mathbf{V}_{\mathbf{GHZ}}(t) \qquad (A.3.8)$$

$$\mathbf{e}_{R2}.\mathbf{e}_{R3}.\mathbf{V}_{\mathbf{GHZ}}(t) = \tfrac{1}{4}.\{(1-i).\mathbf{e}_{L1} + (1+i).\mathbf{e}_{R1}\}.V(t) = \mathbf{e}_{L2}.\mathbf{e}_{L3}.\mathbf{V}_{\mathbf{GHZ}}(t) \qquad (A.3.9)$$

$$\mathbf{e}_{L2}.\mathbf{e}_{R3}.\mathbf{V}_{\mathbf{GHZ}}(t) = \tfrac{1}{4}.\{(1-i).\mathbf{e}_{R1} + (1+i).\mathbf{e}_{L1}\}.V(t) = \mathbf{e}_{R2}.\mathbf{e}_{L3}.\mathbf{V}_{\mathbf{GHZ}}(t) \qquad (A.3.10)$$

Be aware that the expressions above are *not* normalized (they are projected vector components, so it is natural to be not normalized). Other useful expressions are:

$$\mathbf{e}_{R} = \tfrac{1}{2}.\{\mathbf{e}_{+}.(1+i) + \mathbf{e}_{-}.(1-i)\};\ \mathbf{e}_{L} = \tfrac{1}{2}.\{\mathbf{e}_{+}.(1-i) + \mathbf{e}_{-}.(1+i)\} \qquad (A.3.11)$$

$$\mathbf{e}_{+} = \tfrac{1}{2}.\{\mathbf{e}_{R}.(1-i) + \mathbf{e}_{L}.(1+i)\};\ \mathbf{e}_{-} = \tfrac{1}{2}.\{\mathbf{e}_{R}.(1+i) + \mathbf{e}_{L}.(1-i)\} \qquad (A.3.12)$$

With the help of these expressions, the values of all (sixty-four) $N_{uvw}^{ijk}$ ($i,j,k = \pm 1$, $u,v,w = l,r$) can be calculated.

**Acknowledgements.**

This contribution was supported by grant PIP 00484-22 from CONICET (Argentina).

## References.